\documentclass[%twocolumn,
aps,preprint,prc,showpacs,eqsecnum,floatfix,nofootinbib]{revtex4}

\usepackage{amssymb,amsmath}
\usepackage{epsfig}

\newcommand{\be}{\begin{equation}}
\newcommand{\ee}{\end{equation}}

\begin{document}

\title{Nuclei embedded in an electron gas}

\author{Thomas~J.~\surname{B\"urvenich}}
\email{buervenich@fias.uni-frankfurt.de}

\author{Igor~N.~\surname{Mishustin }}
\email{mishustin@fias.uni-frankfurt.de}

\author{Walter~\surname{Greiner}}
\email{greiner@fias.uni-frankfurt.de}

\affiliation{Frankfurt Institute for Advanced Studies, Johann Wolfgang Goethe University,
Max-von-Laue-Str. 1,
60438 Frankfurt am Main,
Germany}

\date{\today}

\begin{abstract}
The properties of nuclei embedded in an electron gas are studied within the
relativistic mean-field approach. These studies are relevant for nuclear properties in astrophysical
environments such as neutron-star crusts and supernova explosions.
The electron gas is treated as a constant background in the Wigner-Seitz cell
approximation.
We investigate the stability of nuclei with respect to $\alpha$ and $\beta$ decay.
Furthermore, the influence of the electronic background on
spontaneous fission of heavy and superheavy nuclei is analyzed. 
We find that the presence of the electrons leads to stabilizing effects
for both $\alpha$ decay and spontaneous fission for high electron
densities. Furthermore, the screening effect shifts the proton dripline
to more proton-rich nuclei, and the stability line with respect to $\beta$-decay
is shifted to more neutron-rich nuclei.
Implications for the creation and survival of very heavy nuclear
systems are discussed.
\end{abstract}

\pacs{21.10.Dr, 21.10.Ft, 21.10.Tg, 23.60.+e, 26.50.+x, }

\maketitle

\section{Introduction}
Nuclear astrophysics lies at the heart of understanding the origin
of the physical world and our mere existence. One of its key questions is 'How
and where did the chemical elements originate?'.
Many challenging questions concerning the processes
and mechanisms
for nucleosynthesis are in the focus of modern research.
In this respect, supernovae and neutron stars
are the most interesting objects to study.
 
The first attempt to explain abundances of chemical elements in the solar
system was made by Geoge Gamow in 1946-48 (see e.g.
Ref. \cite{Alp}). He proposed a mechanism that was based on a continuous building-up 
of chemical elements
by neutron capture in the early universe. Later it became clear that this mechanism
cannot explain the production of heavy elements because of very high specific
entropy of the early universe, S/B$\sim 10^{10}$, where S is the total entropy and B is
the net baryon number. 
The present understanding of the element creation says that 
the lightest elements (up to He, and, partly, Li)
were formed during the first moments of the universe expansion, immediately
after the Big Bang.
Heavier elements up to oxygen are predominantly produced
in thermonuclear reactions in stars like our Sun. Elements up to iron and nickel could be
produced in more massive stars. 
The origin of heavier
elements such as gold and uranium remains in the focus of current research.
It is widely believed that heavy elements were mostly synthesized in the course of
supernova explosions in the so-called r-process (subsequent neutron capture by stable and unstable
nuclei, see a recent review \cite{r-process}).

A type II supernova explosion is one of the most spectacular
events in astrophysics, with huge energy release of about
$10^{53}$ erg or several tens of MeV per nucleon \cite{Bethe}.
When the core of a massive star collapses, it reaches densities
several times larger than the normal nuclear density $\rho_0=0.16$
fm$^{-3}$. The repulsive nucleon-nucleon interaction gives rise to
a bounce-off of matter and formation of a shock wave propagating through the
in-falling stellar material, predominantly Fe.

Hydrodynamical simulations (see e.g. Refs.
\cite{Janka,Thielemann}) show that during the collapse and
subsequent explosion the temperatures $T\approx (0.5\div 10)$ MeV
and baryon densities $\rho_B \approx (10^{-5}\div 2) \rho_0$ can
be reached. Unfortunately, these simulations do not
produce successful explosions yet, even when neutrino heating and
convection effects are included. This means that some important
physics of this process is still missing.

Besides supernova explosions, proto-neutron and neutron stars are interesting objects from a nuclear
physics point of view. They are quite similar to  huge, extremely neutron-rich macroscopic nuclei that
gain additional stability and binding due to the gravitational force.
Neutron stars are very compact objects with a central density of about $\rho = 10^{15} g/cm^3$,
a typical radius of $R = 10$~km and masses up to 2 solar-masses.
Proto-neutron stars are newly-born neutron stars formed in the course of a supernova explosion. They
are somewhat bigger than neutron stars and have temperatures up to 30 MeV.
The regime of interest for our considerations are baryon densities in the range  
$0.001 - 0.5\rho_0$, where very heavy and neutron-rich nuclei may be present
\cite{Lamb,Lattimer,Botvina04,Botvina05}. 
Nuclear pasta phases, i.e., nuclear matter in various geometries such as slabs and parallel plates, reminescent of {\em pasta} \cite{ravenhall83,horowitz06,maruyama05,maruyama06} start at densites slightly above $0.5~\rho_0$. In this
regime the neutrons need to be taken into account.

Since the heaviest elements occuring in nature are Uranium isotopes, there should
have been corresponding extreme conditions for their creation and persistence, in supernova explosions, crusts of
neutron stars, or other sites.
Thus we may wonder if very heavy and superheavy systems can and will be produced under such conditions too.
How are the properties of such nuclear systems altered in dense environments?

The properties of nuclei in astrophysical environments have enjoyed continuous interest
for over more than 30 years.
The seminal work by Negele and Vautherin \cite{negele73} layed out the path for microscopic
studies of nuclei in stellar environments, see e.g. Refs. \cite{Botvina04,magierski01,magierski01-2,douchin98,
sandulescu04,cheng97,maruyama06,toki2000,maruyama05}. 
The investigation of the rich properties of nuclear systems 
under extreme conditions in astrophysical environments is on the rising path.
Since many  astrophysical calculations, from r-process
to dynamical simulations of supernova explosions, depend on theoretical
nuclear input, it is important to understand how nuclei subject
to these special conditions differ from nuclei studied in the laboratory
on earth. 

In this paper we focus on the physics 
of nuclei embedded in a dense electron gas. This investigation
is important by itself as well as for further studies including
the neutron gas.  We consider nuclei
across the perodic chart up to superheavy nuclei embedded in a Wigner-Seitz cell with constant
electron density. 
The presence of the electrons
will effect the location of the 
$\beta$-stability line and the proton dripline as well as decay modes such as $\alpha$-decay and spontaneous fission.

The paper is structured as follows: In Section \ref{framework}, we present the relativistic
mean-field model which is employed in the calculations as well as the concept
of the Wigner-Seitz cell and its concrete implementation for spherical and deformed
nuclear systems. In Section \ref{spherical_calculations}, we  discuss the $\beta$-equilibrium condition
and present results for the $\beta$-stability line and the
proton and neutron drip lines in the presence of electrons.  
The evolution of the $\alpha$-decay mode in heavy nuclei
as a function of the electron Fermi momentum
 is demonstrated in Section \ref{alphadecay}. In Section \ref{deformedcalculations}, deformed
nuclei are considered and spontaneous fission under the influcence of the electron
background is studied. Finally, in Section \ref{conclusions} we conclude and give
an outline of future research directions.
\section{The framework}
\label{framework}

\subsection{The RMF model}

Over the years self-consistent mean-field models have reached high predictive power. They can be applied from
medium-light systems up to superheavy nuclei and to systems
ranging from the proton drip line to the neutron drip line. They are based on the formulation of
an effective nucleon-nucleon interaction meant to be employed in the Hartree or Hartree-Fock
treatment of nuclear systems. The modern way of formulating them is in terms of
an energy functional that can incorporate terms which can not be constructed via a 
two- (or three- or four-) body force.

In this paper we employ the relativistic mean-field (RMF) model \cite{rei89,bender-review}. The effective in-medium nucleon-nucleon
interaction is parametrized via the exchange of several meson fields:
scalar-isoscalar ($\sigma$), vector-isovector ($\omega_\mu$) and vector-isovector ($\vec{\rho}_\mu$).

This model is based on an effective Lagrangian of the form
\begin{eqnarray}
{\mathcal L} &=& \sum_\alpha w_\alpha \bar{\psi}_\alpha (i\gamma_\mu\partial^\mu -m_N)\psi \nonumber \\
& + & 
 \frac{1}{2} \partial_\nu\sigma\partial^\nu\sigma - \frac{1}{2} m^2_\sigma \sigma^2
-\frac{b}{3} \sigma^3 - \frac{c}{4} \sigma^4
- g_\sigma \sum_\alpha w_\alpha\sigma \bar{\psi}_\alpha  \psi_\alpha \nonumber \\
& - &
 \frac{1}{4}
\omega_{\mu\nu}\omega^{\mu\nu} - \frac{1}{2}m^2_\omega \omega^\mu\omega_\mu
- g_\omega \sum_\alpha w_\alpha\omega^\mu \bar{\psi}_\alpha \gamma_\mu \psi_\alpha \nonumber \\
& - &
 \frac{1}{4}
\vec{\rho}_{\mu\nu}\cdot\vec{\rho}^{\mu\nu} - \frac{1}{2} m^2_\rho \vec{\rho}^\mu\cdot\vec{\rho}_\mu
- g_\rho \sum_\alpha w_\alpha\vec{\rho}^\mu \cdot \bar{\psi}_\alpha \gamma_\mu\vec{\tau} \psi_\alpha \nonumber \\
 &-& \frac{1}{4}  F_{\mu\nu} F^{\mu\nu}
- e \sum_\alpha w_\alpha A^\mu \bar{\psi}_\alpha  \gamma_\mu \frac{1 + \tau_3}{2} \psi_\alpha 
%\sum_{i=\sigma, \omega, \rho, \gamma} s_i \big[\frac{1}{2}
%\partial_\nu\hat{\phi}_i^\mu\partial^\nu\phi_{i\mu} - m^2_i \phi^\mu_i\phi_{i\mu})
%- g_i \sum_\alpha \hat{\psi}_\alpha w_\alpha \Gamma_i^\mu\phi_{i\mu}\psi\big]
\label{lagrangian}
\end{eqnarray}
The field tensor for the $\omega$ meson is defined as $\omega_{\mu\nu} = \partial_\mu\omega_\nu
-\partial_\nu\omega_\mu$, and similar definitions hold for the field tensors of the $\rho$
meson and the photon.
The parameters $g_\sigma, g_\omega, g_\rho, m_\sigma, b, c$ are fitted to
experimental data of nuclear ground-state observables. The masses of the $\rho$
and $\omega$ mesons are fixed at the experimental values, since the performance
of the model is not quite sensitive to their values.
As it is written down in the Hartree approximation, the meson fields are treated as classical potentials and
the nucleons are represented by Dirac spinors. 
The isoscalar-scalar $\sigma$ meson delivers the intermediate-range attraction, while the
isoscalar-vector $\omega$ meson is responsible for the short-range repulsion.
The isovector-vector $\rho$ meson couples to the isovector nucleon density and thus
parametrizes the isovector properties of the model. The photon is coupled in
standard fashion. For all boson fields, no (explicit) exchange terms are taken
into account.

It is worth mentioning that these meson fields have only loose correspondence
with the physical meson spectrum. Mean-field models employing contact interactions between
nucleons have a comparable predictive power for nuclear ground-state observables
and excited states \cite{nhm,rmf-pc,rpa}.

The $w_\alpha$ denote occupation probabilities of the nucleon states and originate
from the treatment of pairing. We employ BCS pairing with a density-independent $\delta$-force,
see Ref. \cite{pairing} for details. 
In time-reversal even-even systems, only the time components of the
vector mesons are non-zero. Proton and neutron states are not allowed to mix, hence only
the third components of the isovector-vector $\rho$ field associated with $\tau_3$ survives.

The single-particle equation for the nucleons reads
\begin{equation}
\big[ i \vec{\gamma} \cdot \vec{\partial} + m_N + g_\sigma \sigma + g_\omega \omega^0 \gamma_0
+ g_\rho \rho^0_3 \tau_3 + e A^0 \frac{1-\tau_3}{2}\gamma_0 \big] \psi = \epsilon \gamma_0 \psi
\end{equation}
The binding properties of nuclear matter and nuclei are generated from the strong scalar and
vector fields, $U_S = g_\sigma \sigma \approx -350$~MeV, $U_V = g_\omega \omega^0 \approx +300$~MeV,
which add up to a normal nucleon potential $U_N = U_S + U_V  \approx -50$~MeV. They add up with the same sign to generate
the strong spin-orbit potential in nuclei, which (in the nonrelativistic limit) is given by
\begin{equation}
V_{ls} \propto \frac{d}{dr} (V_S - V_V \gamma_0) \vec{l} \cdot\vec{s}
\end{equation}
This spin-orbit force emerges from the covariant formalism with the right sign and
magnitude without introducing additional parameters. This is an important consequence of
the relativistic description.

In order to account for the electron background, we modify the source term
of the photon field by adding the electron density, i.e.
\begin{equation}
\Delta \phi = -e (\rho_p - \rho_e) = - e \rho_{ch}, 
\label{poisson}
\end{equation}
which is the Poisson equation for the electrostatic potential $\phi \equiv A_0$.
Further down we discuss the concrete implementation in different symmetries.
All other equations remain the same as in vacuum.

Calculations are performed in coordinate space, but the derivatives
are performed as matrix multiplications in Fourier space.
We employ the mean-field parametrization NL3 in these studies  \cite{lalazissis} which delivers
accurate values for nuclear ground-state properties. Furthermore, since the external electron background
couples only electromagnetically to nucleons, no additional parameters need to be introduced,
and no readjustment of the present parameters is needed.

\subsection{The Wigner-Seitz approximation}
\begin{figure*}[t]
\centerline{\epsfxsize=12cm \epsfbox{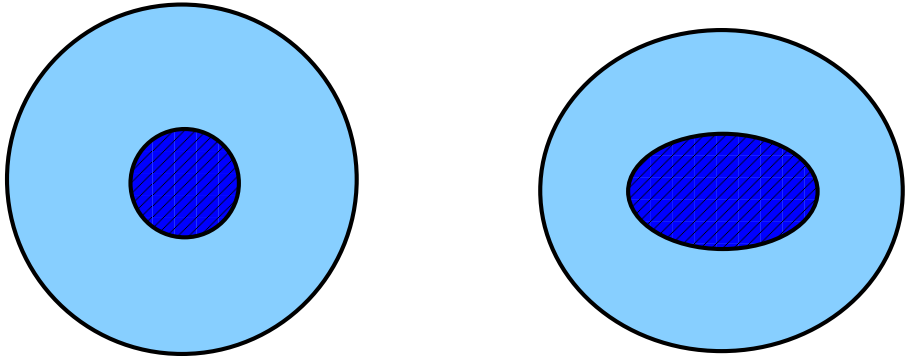}}
\caption{(COLOR ONLINE): Illustration of a Wigner Seitz cell for a spherical nucleus (left)
and an axially deformed nucleus (right). The inner dark filled region 
denotes the charge density of the nucleus, the outer region denotes the extension of the 
electron background surrounding the nucleus. The figure is not to scale.} 
\label{wscell}
\end{figure*}

In this paper we study properties of nuclei embedded in a uniform electron gas at zero temperature.
The calculations are performed within the Wigner-Seitz (WS) approximation by
dividing the system into WS cells, each containing only one nucleus and the number of
electrons equal to the nuclear charge Z, see Fig. \ref{wscell} for an illustration.

The construction of the WS cell is aimed at an approximate and convenient
segmentation of an ensemble of nuclei surrounded by the uniform background of electrons.
Its construction should ensure that the physics contained in the Wigner-Seitz cell is
to an optimally large degree isolated from the surrounding. This requires, for example,
charge neutrality of the cell. However, there could be also other requirements which minimize the electrostatic interactions with neighbouring cells (see below).
Recently the validity of the WS approximation for the inner neutron-star crust has
been discussed, e. g. in  Ref. \cite{baldo06}.

At electron densities considered in this work, the Coulomb potential
due to the nucleus is a small perturbation compared to the Fermi energy
of the electrons. Thus the electrons will only be weakly affected by the presence of the
nucleus which justifies the approximation of constant density.

In the numerical implementation, however, we use a Fermi-type
distribution of the electron density with a smooth boundary. The reason is that on a spherical grid
in coordinate space, the cell radius cannot be fixed exactly (it is limited to
any value that is a multiple of the grid spacing $\Delta r$). A smooth surface
region allows us to realize the electron shell for any grid spacing
and radius. The parameters of the
shell are chosen to satisfy the charge neutrality condition $\int~dV \rho_{ch}(\vec{r}) = 0$, where $\rho_{ch}(\vec{r})$ is the total
charge density. 
Since in the computer code the derivatives are calculated in Fourier space,
the numerical realization of the smooth surface region is not very sensitive
to the chosen grid-spacing.   
This construction of the electron density allows us to
calculate nuclei throughout the nuclear chart without introducing artifical
jumps in the energy caused by a step-like change of the WS cell radius.

\subsubsection{The spherical cell}
\label{WSsph}
For a given nuclear charge Z the WS cell radius $R_{C}$ is uniquely determined by the
charge neutrality requirement
\begin{equation}
\frac{4\pi}{3} R_C^3 \rho_e = Z
\end{equation}
Expressing the electron density in terms of the electron Fermi momentum $k_F$, 
$\rho_e = \frac{1}{3\pi^2} k_f^3,$ one can write the cell radius as
\begin{equation}
R_{C}  
= \Big(\frac{9\pi Z}{4}\Big)^{1/3} \frac{1}{k_F}
\label{wsrad}
\end{equation}
The parametrization with a smooth surface, as used in this work, is given by
\begin{equation}
\rho_e (r) =  \frac{\rho_{e0}}{1 + \exp \frac{(r-r_C)}{a}},
\end{equation}
where $\rho_{e0}$ is the background electron density, $r_C$ is adjusted in order to fullfill charge neutrality, and $a$
is the diffuseness parameter of the distribution, taken to be $0.45$~fm. The
parameter $a$ is chosen such that the charge density is dropping from $90\%$ to $10\%$
of its value at $r=0$ within the diffuseness interval $\sigma = 4.4*a = 2~{\rm fm}$.
We have checked that the results, in particular the differences of nuclear binding energies, are not sensitive  to the choice of $a$. 
The Poisson equation (\ref{poisson}) has to be solved numerically.

While these calculations can be performed quite easily for
high electron densities, for low
electron densities when the WS cell radius becomes quite large, numerical problems arise. For example, for a tin nucleus,
at $k_F = 0.1$~fm$^{-1}$, the WS cell radius is $R_C \approx 70$~fm, while for
$k_F = 0.01$~fm$^{-1}$ it corresponds to $R_C \approx 700$~fm.

An alternative way of calculating the effect of the electron gas is to add the potential of a homogenously
charged sphere of electrons to the electric potential of the proton charge distribution, i.e.,
first calculate the proton electric potential,
\begin{equation}
\Delta \phi_p = - e \rho_p
\label{poisson_e}
\end{equation}
then add the potential due to electrons, 
\begin{equation}
\phi = \phi_p + \phi_{e}
\label{A0}
\end{equation}
where $\phi_p$ is the solution of the Poisson equation for the proton charge density.
The electric potential caused by the uniform background of electrons reads
\begin{eqnarray}
\phi_{e}(r) = \left\{ \begin{array}{l} \frac{-Z e}{4\pi R_{C}} \big[ \frac{3}{2} - \frac{1}{2}
\frac{r^2}{R_{C}^2} \big], \quad r < R_{C} \\
 \frac{-Z e}{4\pi r}, \quad r \geq R_{C}
\end{array} \right.
\label{elec_pot}
\end{eqnarray}

\begin{figure*}[t]
\centerline{\epsfxsize=11cm \epsfbox{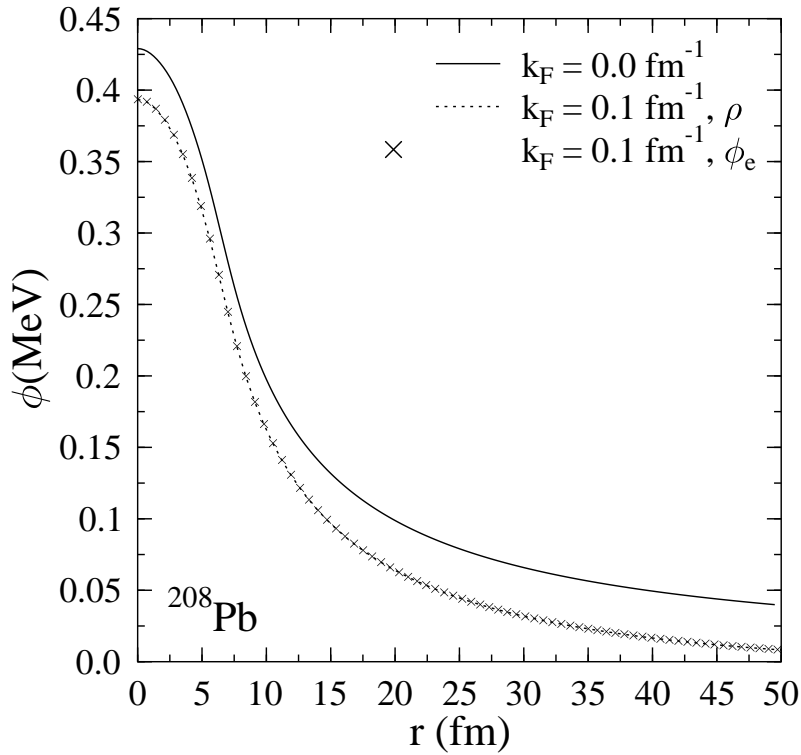}}
\caption{The electrostatic potential $\phi$ shown for the doubly-magic lead nucleus for
the cases of no electrons, for the inclusion of the electron density ($\rho$) and for
the inclusion of the external electron potential ($\phi_e$), respectively, at $k_F = 0.1~$fm$^{-1}$.}
\label{pb_pot}
\end{figure*}

As can be seen from Fig. \ref{pb_pot}, the two different ways of calculating the total electric potential, Eq. (\ref{poisson}) and
Eq. (\ref{A0}), lead to identical electric potentials (apart from the correction due
to the smooth surface region in Eq. (\ref{poisson})) and, therefore, to the same structure of the nucleus. 
This demonstrates consistency and accuracy of the treatment of the electron background
in our numerical calculations.

It is clear that the total energy of the cell should include the interaction energy of the electrons
with the electromagnetic field, which does not appear when the 
electrons are employed as an external potential, Eq. (\ref{A0}).
This interaction energy can be easily calculated for the case when the (spherical) electron and proton
charge distributions are replaced by step-functions:
\begin{equation}
\rho_e(r) = \rho_{e0} ~\theta(R_C - r), \quad \rho_p(r) = \rho_{p0} ~\theta(R_A -r)
\label{theta}
\end{equation}
where $R_C$ and $R_A$ are the cell radius and nuclear radius, respectively. An elementary calculation
yields for the total electrostatic energy of the cell
\begin{equation}
E_{Coul} = \frac{3}{5} \frac{Z^2 e^2}{4\pi R_A} c\big(\frac{\rho_{e0}}{\rho_{p0}}\big), \quad c(x) = 
1 - \frac{3}{2} x^{1/3} + \frac{1}{2} x
\label{e_coulomb_theta}
\end{equation}
This expression has the correct behavior at $x\rightarrow 0$, when it goes to the Coulomb energy of
an isolated nucleus, and at $x\rightarrow 1$, when it gives zero (electron and proton charges
fully compensate each other).
One can see that the Coulomb energy of the nucleus is screened by the electrons. This screening
effect is very significant even at moderate electron densities. For instance, at $x = 10^{-3} ~(k_F \approx 25~{\rm MeV})$ the screening effect is about $15\%$. As we shall see below, this change should lead to
important modifications of the decay channels involving charged particles, such as $\alpha$-decay and
fission.

\begin{figure*}[t]
\centerline{\epsfxsize=11cm \epsfbox{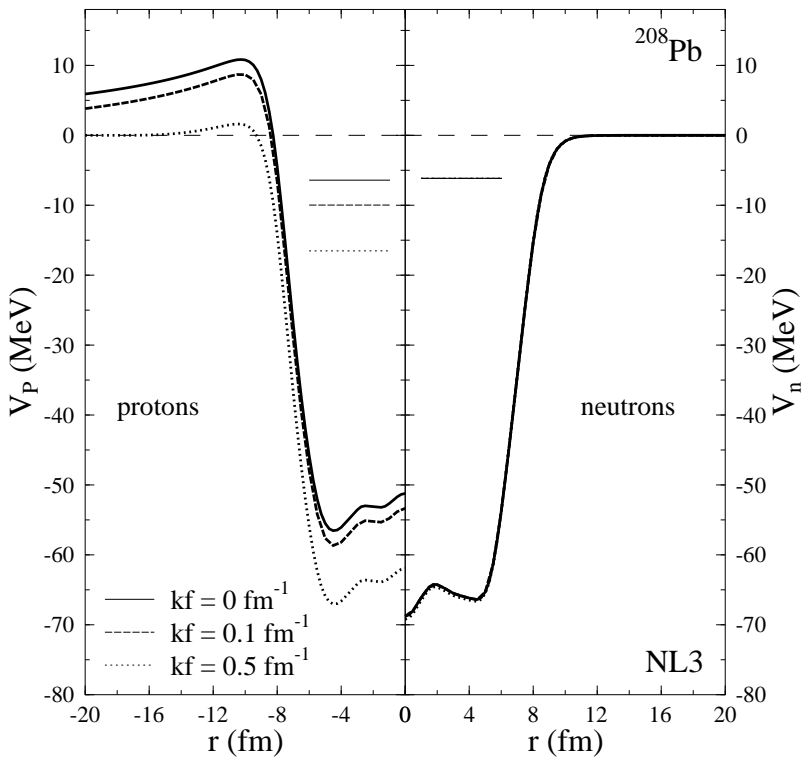}}
\caption{The proton (left) and neutron (right) single-particle potentials in $^{208}$Pb for various
electron Fermi momenta. The proton and neutron Fermi energies are shown as horizontal bars}
\label{pb_pot2}
\end{figure*}

The influence of the spherical electron cloud on the self-consistent proton potential is demonstrated in Fig. \ref{pb_pot2} for $^{208}$Pb.
We note that this nucleus, in the presence of electrons, is not
stable anymore and would undergo electron capture. It is displayed here
to demonstrate the effect of the attractive interaction
between electrons and protons.
The electron background leads to a downward shift of the potential but its structure
remains basically unchanged within the nuclear volume. 
However, the principle difference from the pure Coulombic potential is evident at larger distances where
the electric
potential reaches zero (and zero derivative) at the boundary of the Wigner-Seiz cell.
This is clearly seen in Fig. \ref{pb_pot2} for $k_F = 0.5$~fm$^{-1}$ and the cell radius $R_C = 16.7$~fm. This behavior
is a result of the charge-neutrality condition which makes the electric
field vanish outside the cell. 
We can conclude that at fixed $Z$ and $N$ the nuclear structure, in particular
the single-particle level spacings, remains to good
approximation unaltered by the presence of the electrons. As an example, we note that the proton rms radius in the heavy
nucleus $^{240}$Pu decreases by approximately $0.5 \%$ when adding electrons with $k_F = 0.5 fm^{-1}$. 

\begin{figure*}[t]
\centerline{\epsfxsize=11cm \epsfbox{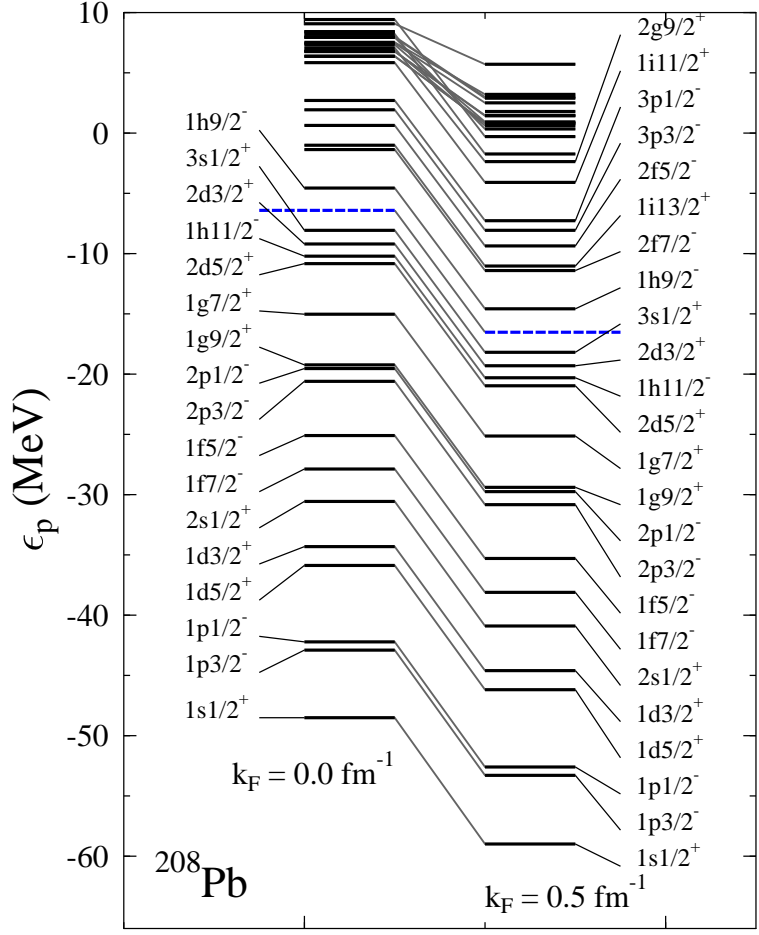}}
\caption{(COLOR ONLINE): The proton single particle levels for $k_F = 0.0~fm^{-1}$ (left) $k_F = 0.5~fm^{-1}$ (right) in $^{208}$Pb. The proton  Fermi energies are shown as horizontal dashed lines}
\label{pb_levels}
\end{figure*}

The Fermi energy of the protons also experiences
a shift downward equal to the amount given by the additional electric potential. The neutron single-particle properties are only
minimally altered, as is expected. The proton single-particle levels of $^{208}$Pb  
are shown in Fig. \ref{pb_levels} for $k_F = 0.0~fm^{-1}$ and $k_F = 0.5~fm^{-1}$. The downward shift of all levels is approximately 10 MeV.
We note that while the level density as such remains the same, the
positioning of the levels with respect to the continuum threshold
has changed. For $k_F = 0.5~fm^{-1}$, the number of proton states
which correspond to bound orbitals has increased, and the density of states close to the continuum has increased. Furthermore, the proton separation
energies have become larger. These changes in the single-particle spectra could affect, for example,
nuclear reactions, in particular electron and neutron capture rates.
\begin{figure*}[t]
\centerline{\epsfxsize=11cm \epsfbox{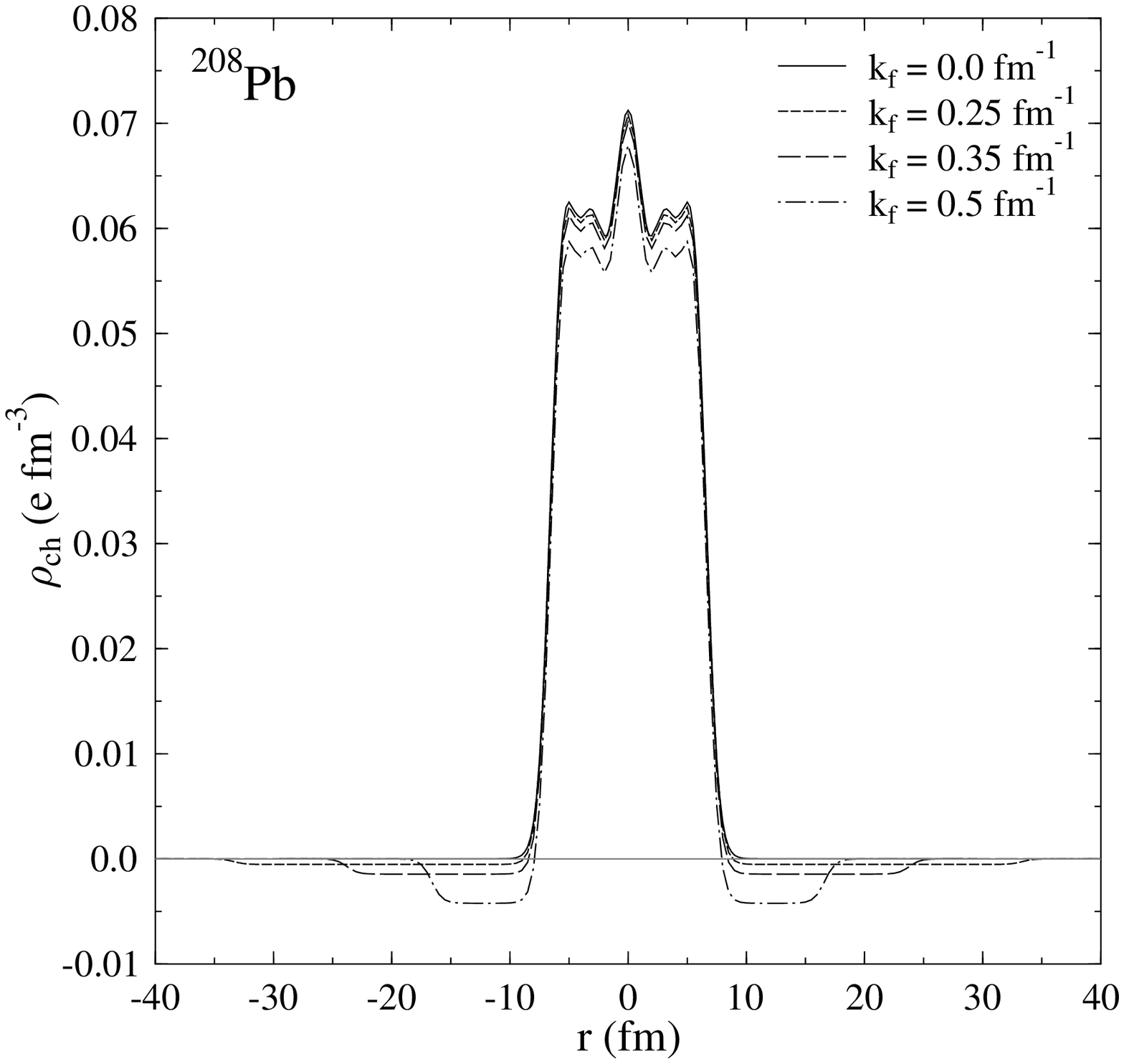}}
\caption{The total charge density in  a cell containing $^{208}$Pb for various
electron Fermi momenta}
\label{pb_chargedens}
\end{figure*}
The total charge density for the doubly-magic lead nucleus embedded in an electron gas in
displayed in Fig. \ref{pb_chargedens}. The negative contribution due to the electrons and the downward
shift in the interior of the nucleus are clearly visible.

\subsubsection{The deformed cell}
For spherical nuclei the spherical shape of the WS cell is the 
obvious choice. 
In this case, the electric field and its derivative vanish on the cell boundary so that
different cells do not experience Coulomb interactions.
However, when we investigate deformed nuclei, there are several possibilities of dealing with such a situation.
One possibility is to employ again a spherical WS cell, with the deformed nucleus sitting in its center. This again corresponds to a constant
electron background that is not affected very much by the presence of the nucleus.
However, due to the deformed proton distribution, the quadrupole moment of the whole cell is nonzero in this case. This means that the neighboring cells will
experience quadrupole-quadrupole interactions.
We think that for the description of deformed nuclei it is more reasonable to use
deformed cells too. Therefore, we consider axially deformed spheroidal
cells with the excentricity determined by the condition of vanishing quadrupole moment.

The quadrupole moment of a given charge distribution is defined as
\begin{equation}
Q_{20} = \frac{1}{2} \sqrt{\frac{5}{4\pi}} \int d^3 x \rho(\vec{r}) (2z^2 - r^2)
\end{equation}
\begin{figure}[t]
\begin{minipage}[h]{13cm}
\centerline{\epsfxsize=13cm \epsfbox{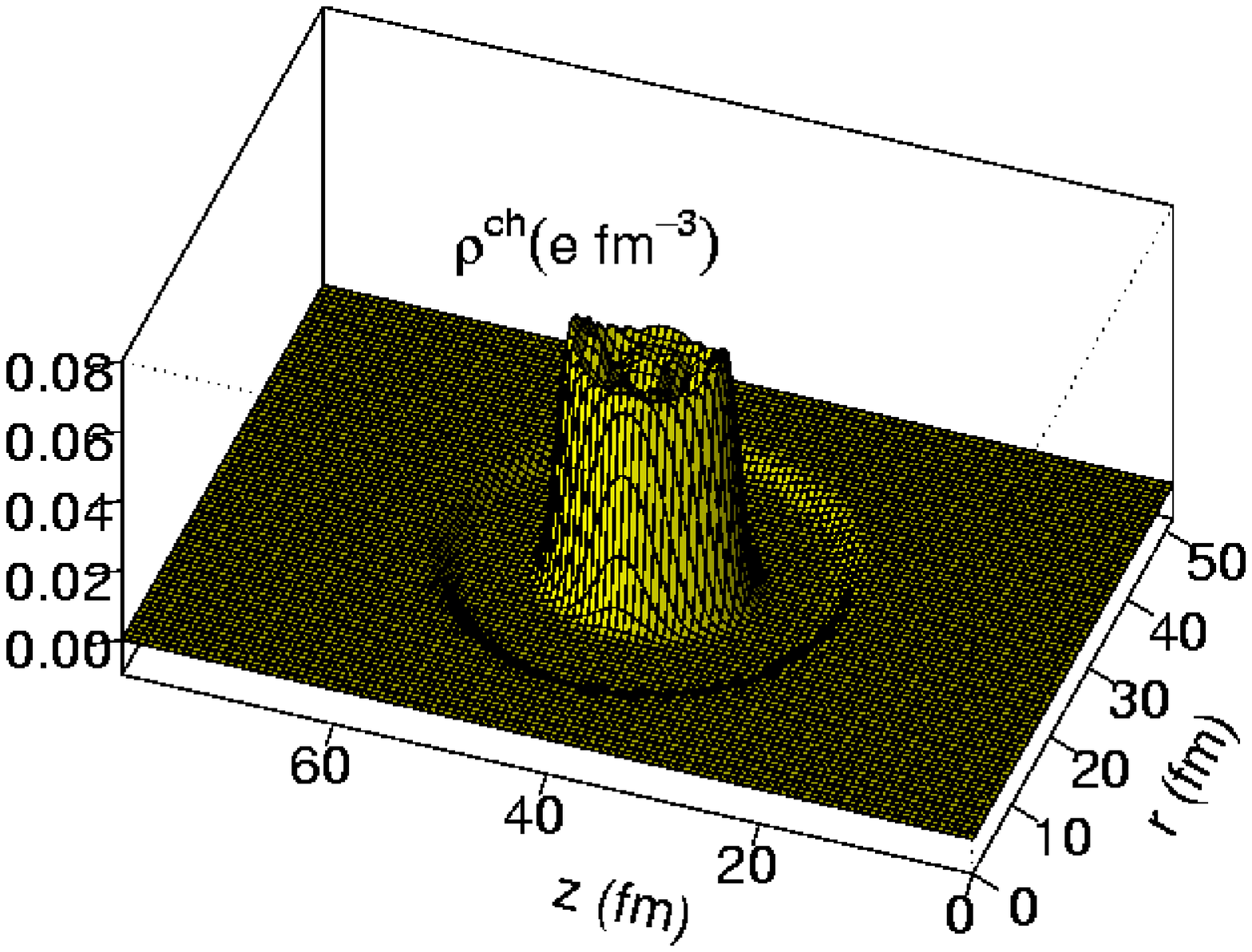}}
\end{minipage}
\begin{minipage}[h]{13cm}
\centerline{\epsfxsize=13cm \epsfbox{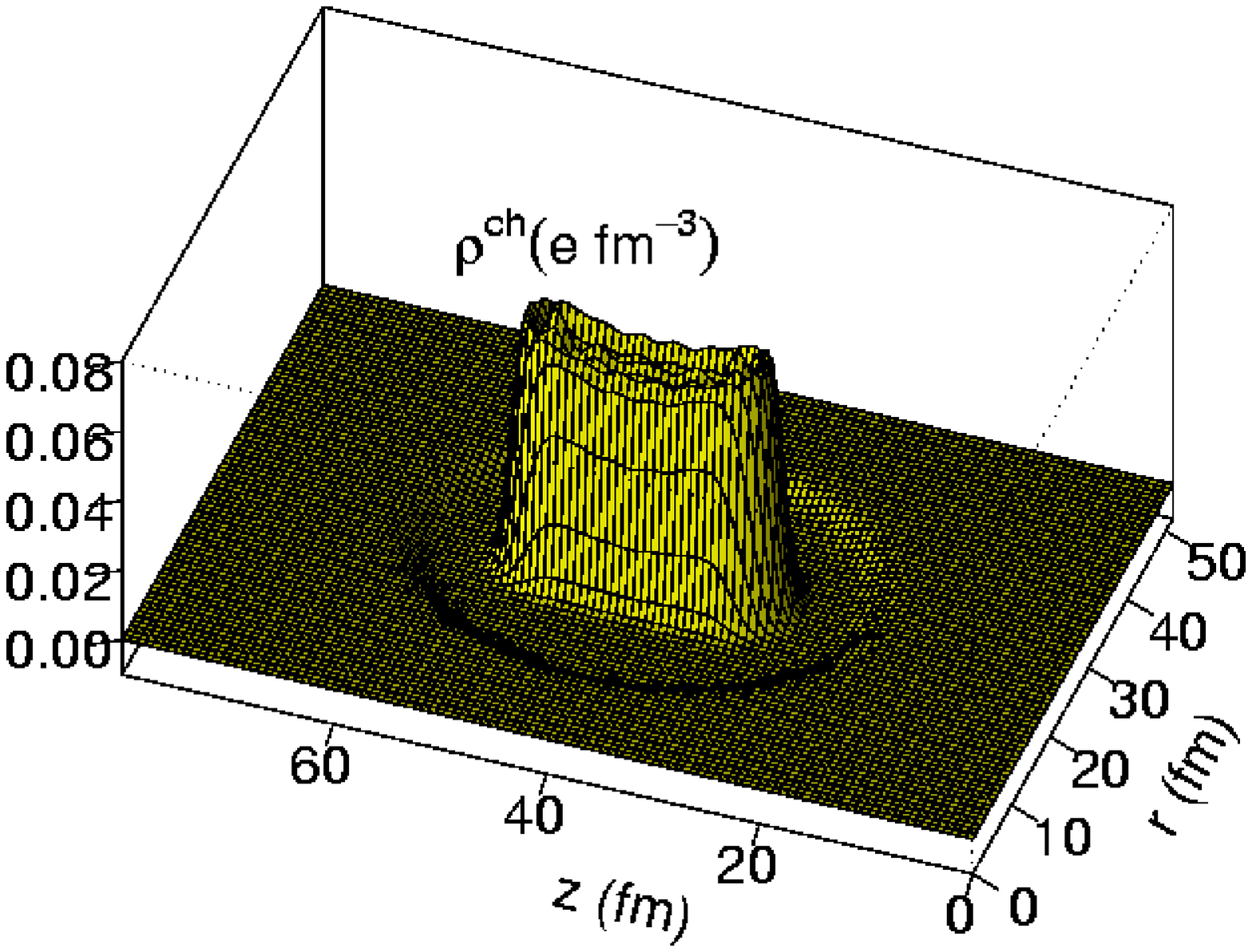}}
\end{minipage}
\caption{(COLOR ONLINE): Total charge density of $^{240}$Pu 
in the ground state (top) and 
at a deformation of $\beta_2 = 1.9$ (bottom) for electron Fermi
momentum $k_F = 0.5 fm^{-1}$}
\label{pu240largebeta}
\end{figure}
\begin{figure}[t]
\begin{minipage}[h]{13cm}
\centerline{\epsfxsize=13cm \epsfbox{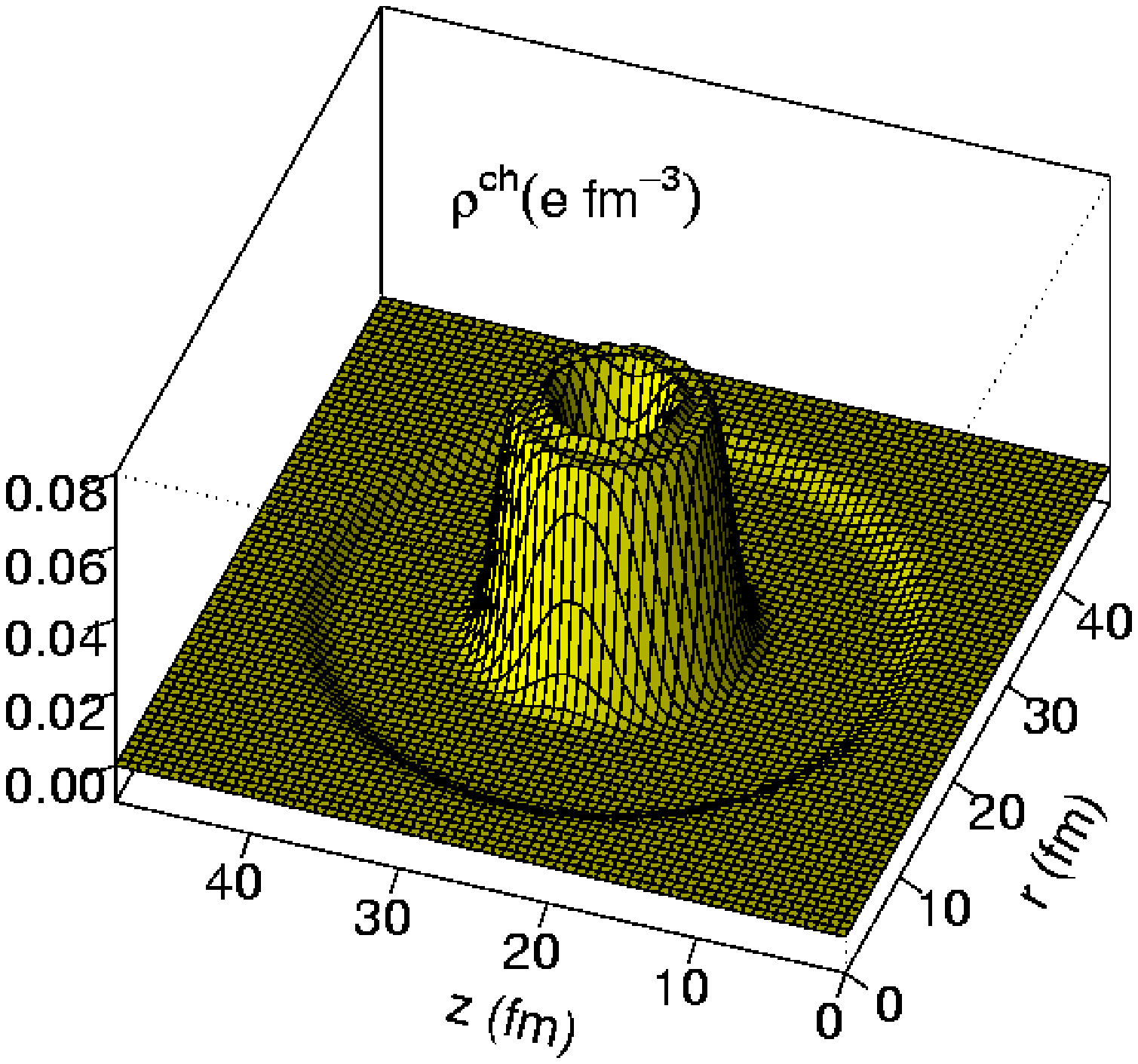}}
\end{minipage}
\begin{minipage}[h]{13cm}
\centerline{\epsfxsize=13cm \epsfbox{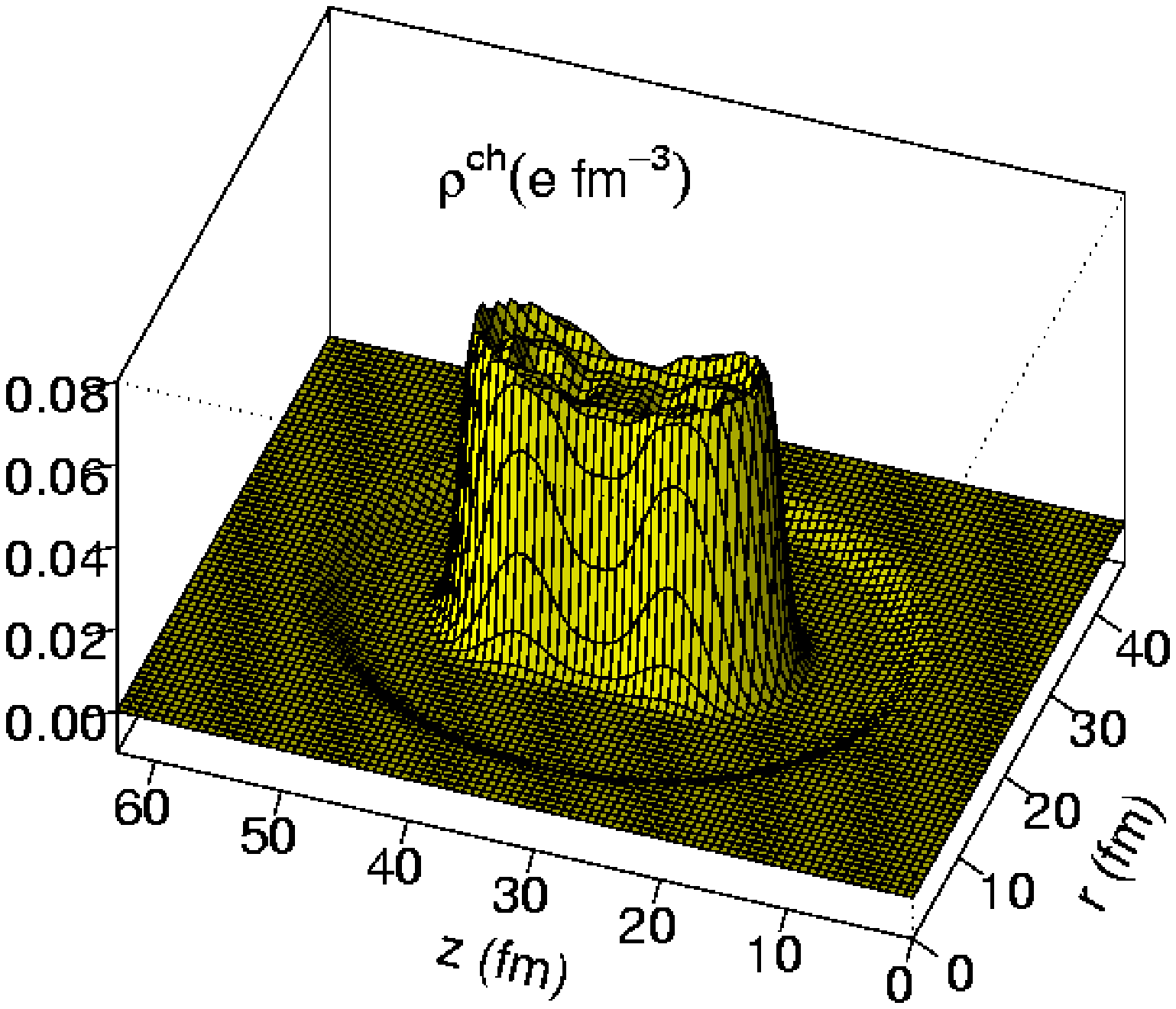}}
\end{minipage}
\caption{(COLOR ONLINE): Total charge density of the superheavy nucleus $^{292}$120 
in the ground state (top) and 
at a deformation of $\beta_2 = 1.5$ (bottom) for electron Fermi
momentum $k_F = 0.5 fm^{-1}$}
\label{120172largebeta}
\end{figure}
In our case $\rho(\vec{r}) = \rho_p(\vec{r}) - \rho_e(\vec{r})$. 
Thus, we adjust the shape of the electron spheroid such that $Q_{20} \equiv 0$.
For the electron Fermi momenta considered in this work the electron charge density
is considerably smaller than the proton charge density at the center of the nucleus, and therefore the 
electron spheroid has larger axes.
At the same excentricity, this would result in a larger negative quadrupole moment
of the electron cloud as compared with the positive quadrupole moment of protons.
Thus, to balance these two contributions, the electron spheroid must have
a smaller excentricity than the proton one.

The realization of the deformed WS cell is illustrated in Fig. \ref{pu240largebeta} for the
ground state and for a largely-deformed state of $^{240}$Pu.
The latter shape is chosen arbitrarily to illustrate the point concerning the
shape of the WS cells.
At the electron Fermi
momentum $k_F = 0.5 fm^{-1}$ the equivalent spherical WS cell radius is 17.5 fm.
In the ground state of Plutonium with a deformation of $\beta_2 = 0.28$, the electron
cloud is almost spherical. 
At the deformation of $\beta_2 = 1.9$, the electron background clearly has a spheroidal
shape. Still, its excentricity is much smaller than the one of the plutonium nucleus which
is much more elongated. This shape corresponds to the nucleus behind the second barrier (see below).
Note that this nucleus will eventually fission through the asymmetric barrier which is energetically
favorable.
However, the spheroidal shape used in our study can be used for axially-symmetric nuclei and,
acoordingly, only for
symmetric fission. Generalizations of this parametrizations would involve more
complicated shapes allowing also for hexadecupole and octupole degrees of freedom.

For the superheavy nucleus $^{292}120$, the corresponding charge densities are
shown in Fig \ref{120172largebeta}. Again, the (spherical) ground-state and
a state at large deformation are displayed. This nucleus exhibits a semi-bubble \cite{hollow1,hollow2} shape
corresponding to a depletion of the baryon in its center that changes with
deformation.

\section{Nuclear chart}
\label{spherical_calculations}
\subsection{$\beta$-stability line}
In this section we analyze how the nuclear $\beta$-stability valley and the proton and neutron drip lines
change with increasing electron Fermi momentum. As shown before, the proton single-particle potential
experiences a constant downward shift. This downward shift of the proton potential and hence
the gap between proton and neutron chemical potentials  is
the reason for the new $\beta$-stability line.  
\begin{figure*}[t]
\centerline{\epsfxsize=15cm \epsfbox{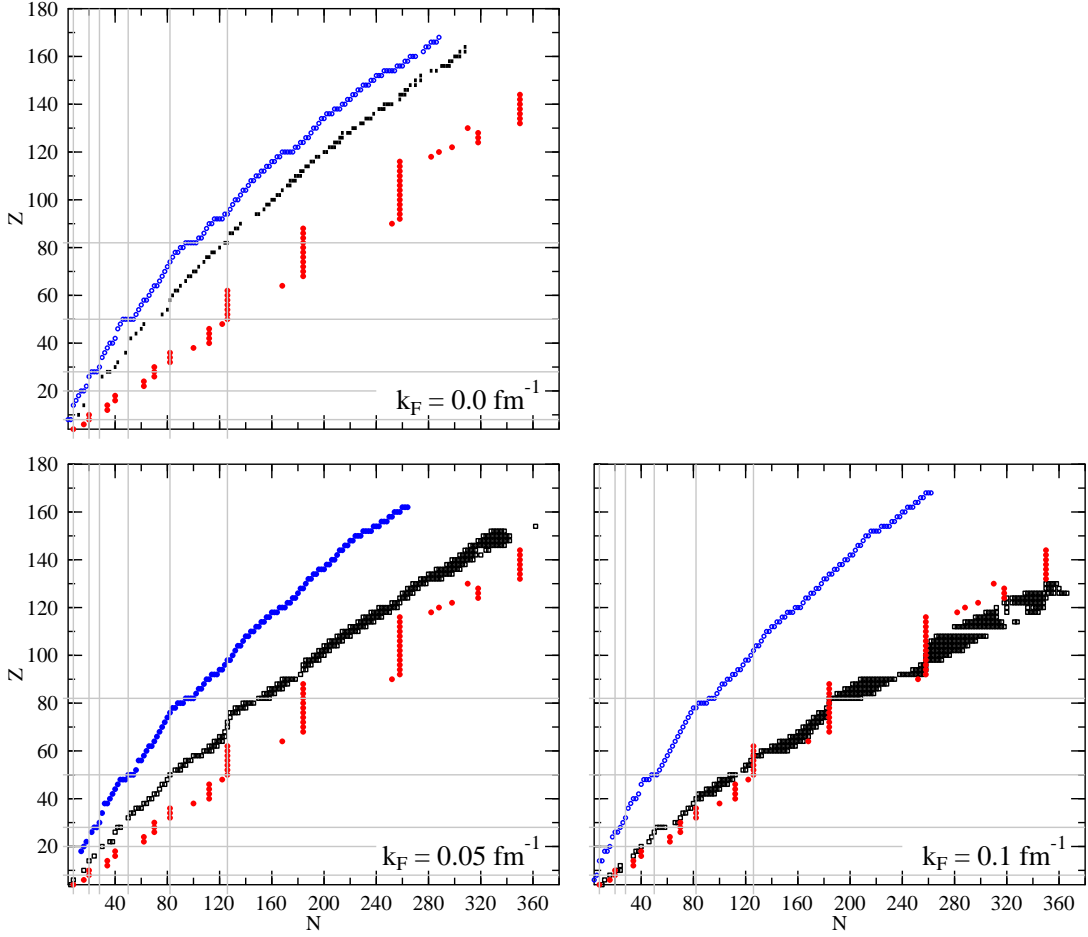}}
\caption{(COLOR ONLINE): The line of $\beta$-stability (line in the middle, black),
the proton drip-line (left-most line, blue) and the two-neutron drip-line (right-most line, red) for various electron fermi momenta.
Magic numbers are indicated by grey lines.}
\label{lines1}
\end{figure*}
In the following, we investigate the possibility of certain decay modes
of nuclei in the electron background.
In particular, electron capture by nuclei leads to a
shift of the stability line to the neutron-rich side.

\label{beta}
In the presence of electrons the equilibrium condition with respect to weak decays
($n \rightarrow p + e^- + \nu_e,~~ p \rightarrow n + e^+ + \bar{\nu}_e$) reads 
\begin{equation}
\mu_n = \mu_p + \mu_e,
\label{betastab}
\end{equation}
where $\mu_n$ and $\mu_p$ are neutron and proton chemical potentials calculated
within the framework of the RMF model, i.e.
\begin{equation}
\mu_N = \sqrt{(m_N+g_\sigma \sigma)^2 + p_{FN}^2} + g_\omega \omega^0 + g_\rho \rho^0\tau_3, \quad
N= n, p
\end{equation}
The electron chemical potential is simply given by 
\begin{equation}
\mu_e = \sqrt{k_F^2 + m_e^2} 
\end{equation}
where the contribution of the electrostatic potential has been neglected since
it cancels out with the corresponding contribution from the proton in Eq. (\ref{betastab}).

If $\mu_n > \mu_p + \mu_e$ holds, neutrons can decay into available proton states
with the electron going on the top of the Fermi distribution.
On the other hand, if $\mu_n < \mu_p + \mu_e$, protons can capture electrons 
from the Fermi distribution and 
occupy neutron states. Only when the condition in Eq. (\ref{betastab}) is fulfilled,
nuclear systems are stable with respect to these decay modes. Note that this relation
becomes the well-known condition for $\beta$-stability in vacuum, $\mu_n \approx \mu_p$,
for $k_F = 0$.

For zero temperature,
the proton and neutron chemical potentials coincide with their Fermi energies.
We note that this correspondence is not uniquely defined anymore
for systems with pairing, where the effective Fermi energy is determined to
yield $\langle \hat{N} \rangle = A$, but can lie somewhere between the last level contributing
with nonzero occupancy and the next one. 
Thus, in order to obtain the (N,Z) dependence of the stability line, instead of Eq. (\ref{betastab}) we employ
the criterion 
\begin{equation}
|\mu_e - (\mu_n-\mu_p)| < \Delta,
\end{equation}
where $\Delta = 1$~MeV.

In Fig. \ref{lines1}, the $\beta$-stability lines  are plotted for various values of $k_F$. In the same plot we show also
the proton and neutron drip lines (see below).
As $k_F$ increases, the line of $\beta$-stability is shifted more and more to the neutron-rich side,
i.e. the presence of electrons stabilizes
neutron-rich nuclei.  This shift is so
strong already for $k_F = 0.1 fm^{-1}$ that $\beta$-stability is reached only in the region of
the two-neutron drip line.  This  means that the $\beta$-equilibrium is reached only when free
neutrons appear in the system.

\subsection{Proton- and neutron drip lines}
\label{driplines}

The neutron and proton drip lines are defined by the conditions $\mu_p = m_p$ and $\mu_n = m_n$,
respectively. 
It is clear that these nuclei are unstable with respect to weak
decays so that the condition of $\beta$-equilibrium, Eq. (\ref{betastab}), does not hold.
In finite systems the dip lines can be
defined in terms of separartion energies
$S_p = B(N,Z) - B(N,Z-1)$ and $S_n = B(N,Z) - B(N-1,Z)$.
Since in this work we are only calculating even-even system, we will
consider instead two-proton and two-neutron drip lines. 
The two-proton drip-line is defined as the position where the two-proton separation
energy, defined as
\begin{equation}
S_{2p}(N,Z) = B(N,Z) - B(N, Z-2),
\end{equation}

goes from positive to negative values. As an approximate relation,
the single-proton separation energies $S_p (N,Z)$ are related to
the two-proton separation energies by $S_{2p} (N,Z) \approx 2\cdot S_{p} (N,Z)$.
\begin{figure*}[t]
\centerline{\epsfxsize=12cm \epsfbox{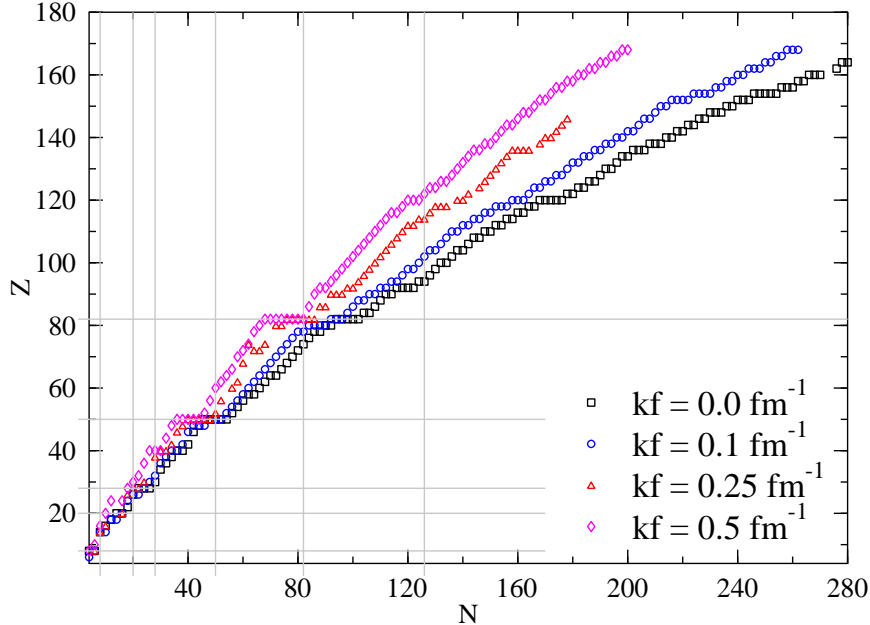}}
\caption{(COLOR ONLINE): The proton drip-line for various electron Fermi momenta. Magic numbers are indicated by grey lines.}
\label{lines2}
\end{figure*}
The neutron drip line is the same for all cases shown in Fig. \ref{lines1} since
its position is not affected by the presence of the electrons. This follows from the fact that
the neutron single-particle potential is not affected by the changes in the proton potential.
Thus the neutron single-particle states are (almost completely) identical in all cases and
the neutron drip line does not change. 
However, the results on the neutron drip line need to be taken with a grain of salt.
Firstly, the calculations here are performed in spherical symmetry and deformation
effects will wash out some of the strong shell effects appearing in these
calculations. Secondly, we employ BCS pairing which overestimates the
coupling to continuum states and the pairing contribution of the loosely bound
orbitals. For nuclei close to or at the neutron drip line, full Hartree-Fock-Bogolyubov
calculations should be performed. However, since here we are not
interested in the precise details of the position of the neutron drip line, the BCS
calculations can give us useful information on the relation between
the line of $\beta$ stability and the neutron drip region.

The evolution of the proton drip-line as a function of the Fermi momentum is -- in addition to Fig. \ref{lines1} -- displayed separately in
Fig. \ref{lines2}. As expected, this drip line shifts to more proton-rich nuclei as
the electron Fermi momentum increases due to the additional binding of protons
produced by the electron background. It is interesting that for $k_F =0.5 fm^{-1}$, the proton drip line is very close
to the $N = Z$ condition. We can summarize our findings by stating that the region of
nuclei between the conditions of proton drip and $\beta$ stability has increased.

\section{Decay modes}
\label{decaymodes}

\subsection{$\alpha$-decay}
\label{alphadecay}
\begin{figure*}[t]
\centerline{\epsfxsize=10cm \epsfbox{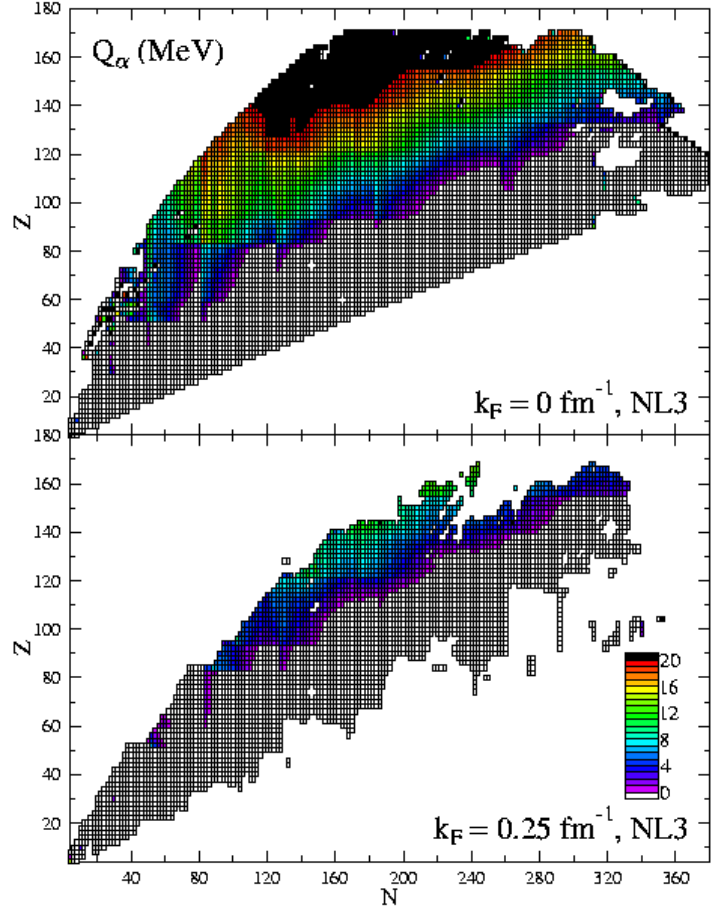}}
\caption{(COLOR ONLINE): $Q_\alpha$-values for $k_F = 0$~fm$^{-1}$ (top) and $k_F = 0.25$~fm$^{-1}$ (bottom).
The lower boundary corresponds to $Q_\alpha = 0$. Nuclei for which
no convergent solution has been achieved have been left out.}
\label{qalphavalues}
\end{figure*}
\begin{figure*}[t]
\centerline{\epsfxsize=10cm \epsfbox{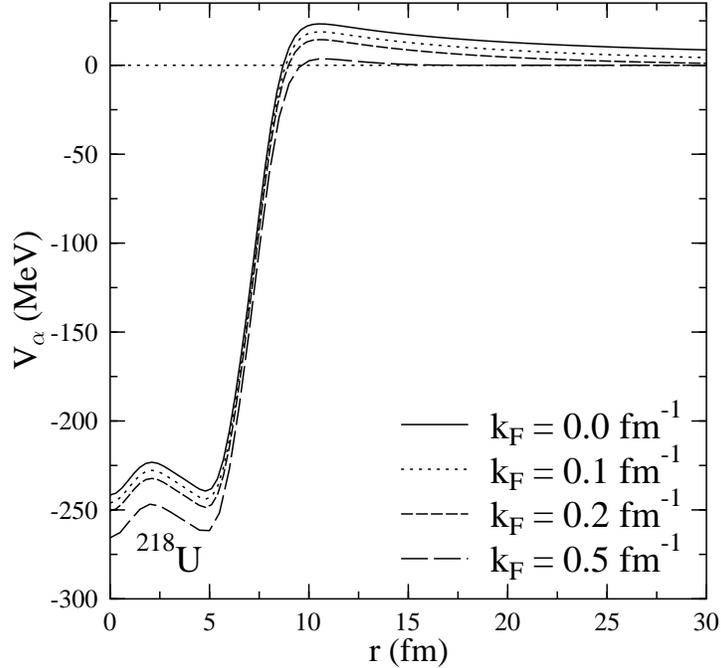}}
\caption{$\alpha$-potential for $\alpha$-decay of the nucleus $^{218}$U. }
\label{alpha_pot}
\end{figure*}
\begin{figure*}[t]
\centerline{\epsfxsize=16cm \epsfbox{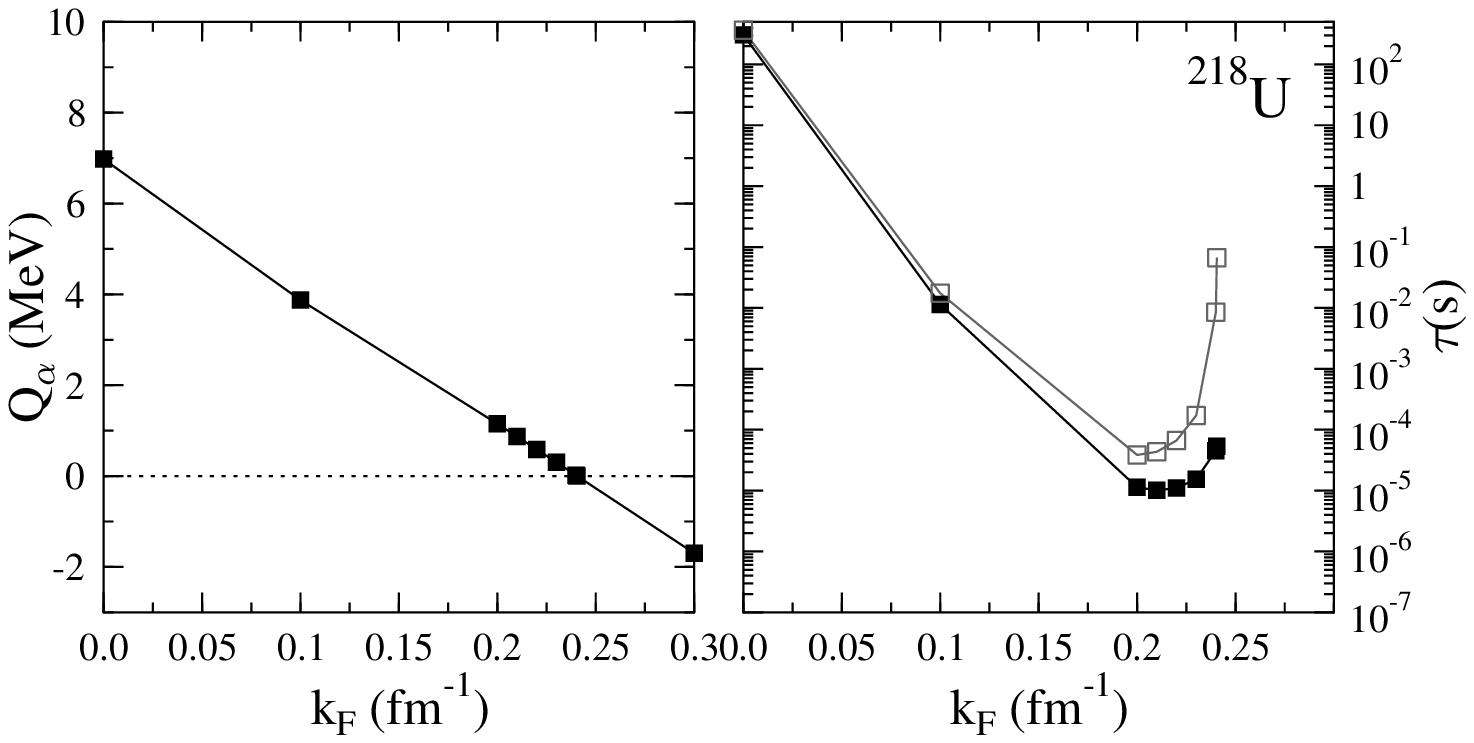}}
\caption{The evolution of $Q_{\alpha}$ values (left) and $\alpha$ decay half-lifes (right) for the nucleus $^{218}$U as a function
of the electron Fermi momentum. In the right panel,
the full squares correspond to calculations within the model of Ref. \protect \cite{zupre06}, and open 
squares give results for life times multiplied with the prefactor of Eq. (\ref{prefactor}).
The lines are drawn to guide the eye}
\label{alpha_life}
\end{figure*}
In this section we investigate the influence of the electron background on $\alpha$-decay of nuclei.
The $Q_\alpha$ value of the reaction is defined as
\begin{equation}
Q_\alpha(N,Z) = B(N,Z) - B(N-2, Z-2) - B(2, 2)
\label{eq_qalpha}
\end{equation}
and corresponds approximately, neglecting nuclear recoil, to the kinetic energy of the $\alpha$ particle leaving
the nucleus. In the considered environment, the binding energies of
mother and daughter nuclei as well as the binding energy of the $\alpha$-particle are
increased due to the attractive electromagnetic interaction of protons with the electron background.
Fig. \ref{qalphavalues} demonstrates this effect.
One can see that the overall trend is that the $Q_\alpha$ values decrease.
Also, the boundary for which $Q_\alpha = 0$  shifts to more proton-rich nuclei.
For $Q_\alpha < 0$, $\alpha$-decay is not possible anymore and these systems
are completely stabilized with respect to this decay mode.

The calculation of the $\alpha$-decay lifetimes in the presence of electrons is not completely trivial.
There are two competing effects. As seen above, the $Q_\alpha$ value is lowered 
due to the increased nuclear binding caused by electrons, which alone (for the same barrier) could increase the lifetime. We have also seen that the presence of the electrons leads to screening and 
modification of the Coulomb potential, see Eq. (\ref{e_coulomb_theta}).
Therefore, we need also to take into account the change of the barrier through which the
$\alpha$ particle has to penetrate. To quantify this effect we use a simple one-parameter model of Ref. \cite{zupre06} which expresses
the $\alpha$-particle potential
 $V_\alpha$  as
\begin{equation}
V_\alpha = 2 ~V_p + 2 ~ V_n,
\end{equation}
where $V_p$ and $V_n$ are, respectively,  proton and neutron single-particle potentials. 
The $\alpha$ particle is treated as a boson.
The half-life is written as $\tau_{1/2} = \ln 2 / \lambda$, where the decay constant $\lambda$
is parametrized as $\lambda = c ~ P$. Here 
 the pre-formation factor and the knocking
frequency in the Gamov picture are absorbed into one parameter $c$, which is adjusted
 to known data \footnote{A model variant that computes the
knocking frequency numerically \protect\cite{zupre06} yields quite similar results.}.  
The probability
for transmission through the barrier, $P = e^S$, is calculated
within the WKB approximation
\begin{equation}
S = -2 \int_{R_1}^{R_2} dr \sqrt{\frac{2\mu[V_\alpha(r) - Q_\alpha]}{\hbar^2}},
\label{wkb-formula}
\end{equation}
where $R_1$ and $R_2$ are the turning points of the barrier.

As demonstrated in Ref. \cite{zupre06},
this model performs better than the 4-parameter Viola-Seaborg systematics and additionally
incorporates isovector trends, i.e. the $\alpha$ lifetime for a nucleus with given
charge number possesses a dependence on the neutron number due to the 
self-consistent interrelation between proton and neutron single-particle
potentials.

We employ this model here also for finite electron density where the potentials and the
$Q_\alpha$ values are taken from the RMF calculations in the electron
background. As an example we present results for the nucleus $^{218}$U
with the magic neutron number $N=126$. We have checked in axially-symmetric calculations that 
both this nucleus as well as its daughter are spherical, thus no deformation
effects enter the results on $\alpha$-decay.

As seen in Fig. \ref{alpha_pot},
the $\alpha$-particle potential is lowered as a function of increasing electron Fermi momentum, making it easier for the $\alpha$ particle
to escape the nucleus. Together with the $Q_\alpha$ value this leads to an overall decreasing of $\alpha$ half lives, see Fig \ref{alpha_life}.
This holds true at low electron densities.
However, this trend changes in the vicinity of the point where the $Q_\alpha$ values turn negative at some electron Fermi momentum $k_F$.
At 
$k_F > k_F^{crit}$, $\alpha$-decay becomes forbidden and nuclei are stable with respect
to this decay mode. In the Uranium isotope discussed here, $k_F^{crit} = 0.24~fm^{-1}$.
However, the potential barrier
experienced by the $\alpha$ particle at $k_F^{crit} = 0.24~fm^{-1}$ is still $\approx 13$~ MeV high. 
If the barrier would result from a pure nuclear charge, the lifetime
for $Q_\alpha \approx 0$ would increase by a huge amount compared to situations with
a finite $Q_\alpha$ value. The reason is that in vacuum the Coulomb barrier
goes to zero asymptotically as $1/r$, thus at $Q_\alpha\approx 0$ the integration distance for the
WKB formula is very large. 
By using a Taylor expansion of the Gamov factor for a $1/r$ potential one can find the approximate
relation $\ln \tau_{1/2} \approx Q^{-1/2}_\alpha$, which leads to an infinite lifetime at $Q_\alpha \rightarrow 0$.
In the case of electron background, however, 
the Coulomb potential is fully screened for distances larger than 
the Wigner-Seitz radius, which decreases with increasing electron Fermi momentum.
Thus, this approximate relation is not applicable anymore.
We expect that the WKB-based calculation underestimates the upwards trend shown in Fig.
\ref{alpha_life} (right), since close to the threshold this method is not accurate \cite{wkb}.

This situation can be illustrated by a text-book 1-D tunneling problem. A particle coming from 
the negative $x$ direction encounters a rectangular barrier of height $V$ between $x = 0$ and
$x = d$. It can be shown \cite{mayer} that the transmission coefficient $T$ is given by
\begin{equation}
T = \Big[ 1 + \frac{V^2}{V^2 - (2E-V)^2} \sinh^2\big(\frac{\sqrt{2m(V-E)}d}{\hbar}\big) \Big]^{-1}
\label{T}
\end{equation}
where $E$ and $m$ are the energy and the mass of the tunneling particle, respectively.
For typical cases of $E \approx V/2$ and $\sqrt{2m(V-E)}d/\hbar >> 1$ (corresponding to
a broad barrier compared to the de Broglie wave length of the tunneling particle),
we have $\sinh(\sqrt{2m(V-E)}d/\hbar) \approx (1/4)\exp(2\sqrt{2m(V-E)}d/\hbar) >> 1$.
Since the pre-exponential is of order 1, we can neglect the first term in brackets of Eq. (\ref{T}) and
obtain 
\begin{equation}
T \approx \exp[-2/\hbar\sqrt{2m(V-E)d}]
\label{Texp}
\end{equation}
This is exactly what follows from the WKB approximation 
of Eq. (\ref{wkb-formula}).
For  electron Fermi momenta close to $k_F^{crit}$, however, we have $E \rightarrow 0$,
and hence the prefactor diverges, i.e., 
\begin{equation}
\frac{V^2}{V^2 - (2E-V)^2} \approx \frac{V}{4E} \rightarrow \infty, \quad E \rightarrow 0
\label{prefactor}
\end{equation}
This gives additional suppression of the tunneling probability as compared to the one obtained in the WKB approximation.
Thus, the trend to longer lifetimes can be expected to set in for even 
smaller values of $k_F$ compared to our 
calculations using the WKB approximation (Fig. \ref{alpha_life}, full squares).
This means that there
is a $k_F^{min}$ for which the lifetime reaches its minimum, and for $k_F > k_F^{min}$
the lifetimes increases until the nucleus becomes stable with respect
to this decay mode for $k_F \geq k_F^{crit}$.
To illustrate the trend we have made estimates of the lifetimes obtained by multiplying
Eq. (\ref{Texp}) with the prefactor of Eq. (\ref{prefactor})
for $E = Q_\alpha$. 
Results are shown in the right part of Fig. \ref{alpha_life} with open squares. 
In this case the reversal of the downward trend in half lives occurs  earlier and more dramatically
compared to the previous calculations.
For the electron Fermi momentum $k_F = 0.1$~fm$^{-1}$ the prefactor has the value $1.53$,
while for $k_F = 0.2405$~fm$^{-1}$ it reaches the value $1255$. A further modification
would arise from the fact that the exponential function in Eq. (\ref{T}) would have to be
replaced by the $\sinh$ function of Eq. (\ref{Texp}) which could yield even more
stabilization with respect to $\alpha$ decay.

The linear decrease of the $Q_\alpha$ values shown in Fig. \ref{alpha_life}
can be understood if we consider the electric potential caused by
a spherical electron cloud given in Eq. \ref{elec_pot}. The potential energy $V_{pot}$ of a point charge located 
in the middle of the cloud is inversely proportional to the WS cell radius $R_C$. Taking into account that $R_C \propto 1/k_F$, Eq. \ref{wsrad}, we obtain
$V_{pot} \propto k_F$, i.e., a point-like nucleus gains potential energy proportional
to $k_F$. If $Z$ denotes the charge number of the mother nucleus and $Z_1$ and $Z_2$ denote
the charge numbers of the daugther products ($Z_1=2$ and $Z_2 = Z-2$ for $\alpha$ decay),
The change in the $Q$ value stemming from the additional binding due to
the electron cloud is given by
\begin{equation}
\delta Q = \frac{-3}{8\pi}\big(\frac{4}{9\pi}\big)^{1/3} e^2 k_F \big[Z^{5/3}-Z_1^{5/3}-Z_2^{5/3}\big]
\end{equation}
Thus the $Q_{\alpha}$ value which depends on
the energies of the mother and daughter nuclei, decreases linearly
with $k_F$. We also see that since $Q \propto Z^{5/3}$ the largest contribution is coming from the mother
nucleus.

\subsection{Spontaneous fission}
\label{deformedcalculations}
\subsubsection{Deformed calculations}
%\label{deformed_calculations}
%
In order to investigate the influence of the electron background on the fission barrier of
heavy and superheavy nuclei, we compute
the energy of the system as a function of deformation.
The cuts through the potential energy surface (PES) are calculated in axial
symmetry for reflection-symmetric shapes using a constraint on the
total quadrupole moment $Q_{20}$ of the nucleus. This is achieved by adding
$-\lambda \,\hat{Q}_{20}$ to the Hamilton operator and minimizing
$\langle\hat{H} - \lambda \hat{Q}_{20} \rangle$. All other multipole moments
that are allowed by the symmetry of the calculation are not constrained and
adjust themselves  to the solution of minimal energy.
So, in contrast to the macroscopic-microscopic approach \cite{mol95}, we are not
operating in a limited deformation space. On the other hand, we are not
necessarily exactly following the gradient in the multidimensional
PES which would only be the case if the fission valley would be parallel
to the $Q_{20}$ direction \cite{floc73}.

At each quadrupole deformation along the fission barrier, the electron clouds
excentricity and spatial extension is iteratively adjusted to match two constraints:
1) the total charge of the deformed WS cell has to be zero, and 2) the quadrupole moment
of its charge distribution (including contributions from both protons and electrons)
should vanish. These two constraints can be realized by varying two parameters characterizing
the electron cloud, for example the lengths of long half-axis and the excentricity.
As discussed above, these conditions minimize the mutual interactions of WS cells.

\subsubsection{Fission barriers}
\label{fissionbarriers}
In fission studies an important role is played by the energy of the system as a function of the deformation parameter $\beta_2$.
Our analysis shows that the shape of the electron cloud has some effect on the 
energy. While the spherical cloud does not produce any changes, the deformed cloud leads to visible
effects. 

Important key quantities related to fission are the width and the height of the fission barrier.
The influence of the electrons on the fission barrier is a macroscopic effect resulting from interaction
of the protons with the Coulomb potential produced by the electrons.
The electrostatic repulsion between the protons is
weakened by the presence of the negative charge background. This change leads to the
increase of barrier height and isomer energy. Physically speaking, the
electron background tends to slightly stabilize the system
with respect to symmetric spontaneous fission and to increase
the excitation energy of the isomeric state.
We expect that asymmetric fission will be altered in a similar way, although our consideration
does not allow octupole shapes.
We note that the potential energy surface is determined by the dependence of the energy
on the deformation. Especially for superheavy systems, the liquid drop barrier
vanishes, and their stabilization results from shell effects.

\begin{figure*}[t]
\centerline{\epsfxsize=12cm \epsfbox{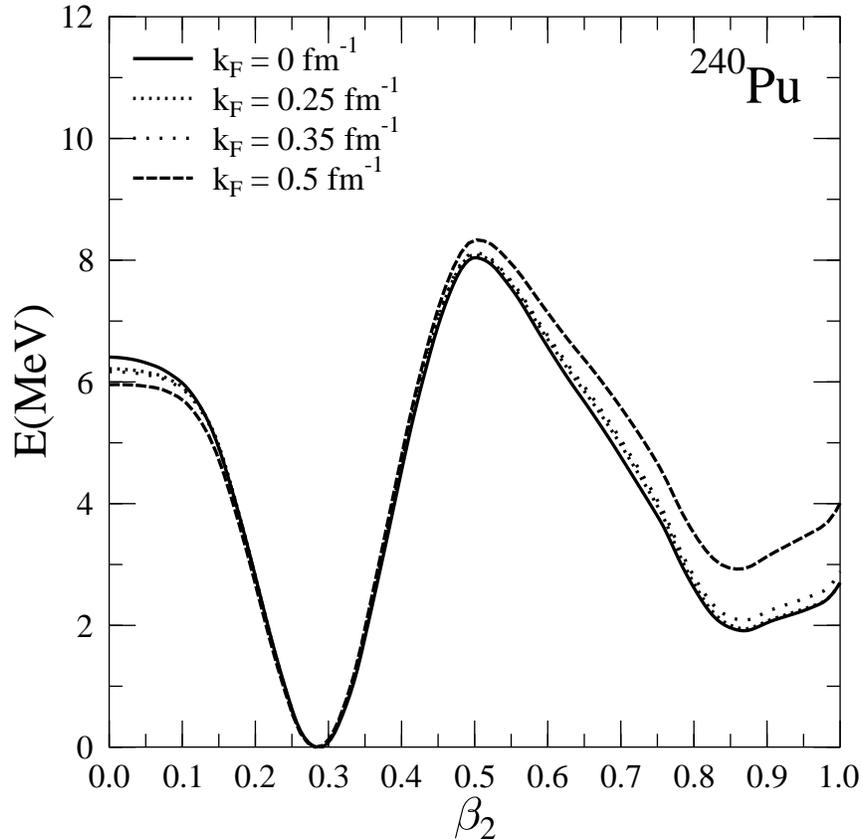}}
\caption{Fission barrier of $^{240}$Pu  for various electron densities in spheroidal configurations.}
\label{pu240_kf}
\end{figure*}
The trends discussed above are clearly seen in
Fig. \ref{pu240_kf} which displays the fission barrier of Plutonium for different choices of the electron
density in the spheroidal configuration. The inner barrier increases with increasing 
$k_F$ leading to a stabilization effect towards spontaneous fission. Furtheron, the properties
of the shape isomer (second minimum) are altered due to the electrons background too. For $k_F = 0.5~fm$ we see an increase of the inner fission barrier by approx. $0.3$ MeV and
the energy of the isomeric state increases by approximately $1$ MeV. Note that the ground-state binding energy of $^{240}$Pu with NL3
is 1814 MeV, the presence of the electrons increases it to the value of
2248 MeV (the latter does not include the kinetic 
energy of the electrons).
\begin{figure*}[t]
\centerline{\epsfxsize=12cm \epsfbox{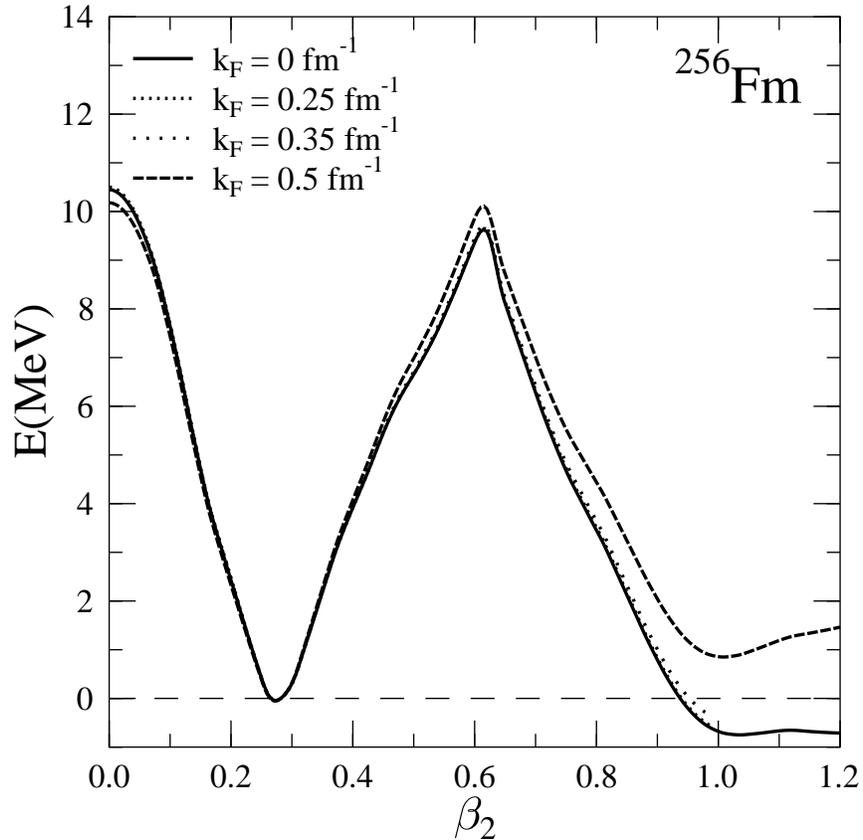}}
\caption{Fission barrier of $^{256}$Fm in for various electron densities in spheroidal configurations.}
\label{fm256_kf}
\end{figure*}
As another example, we show the influence of the electron background on the fission barrier
of $^{256}$Fm in Fig. \ref{fm256_kf}. The presence of the electrons leads to a slight increase of
the first barrier. The effect is more dramatic for the isomeric state, where an increase
in energy of 2 MeV is found at $k_F = 0.5 fm^{-1}$. Note that the low energy of the second minimum, which is below
the ground state, is a (unrealistic) prediction which has also been found in a systematic
study for heavy and superheavy nuclei \cite{shebarrier}.

\begin{figure*}[t]
\centerline{\epsfxsize=12cm \epsfbox{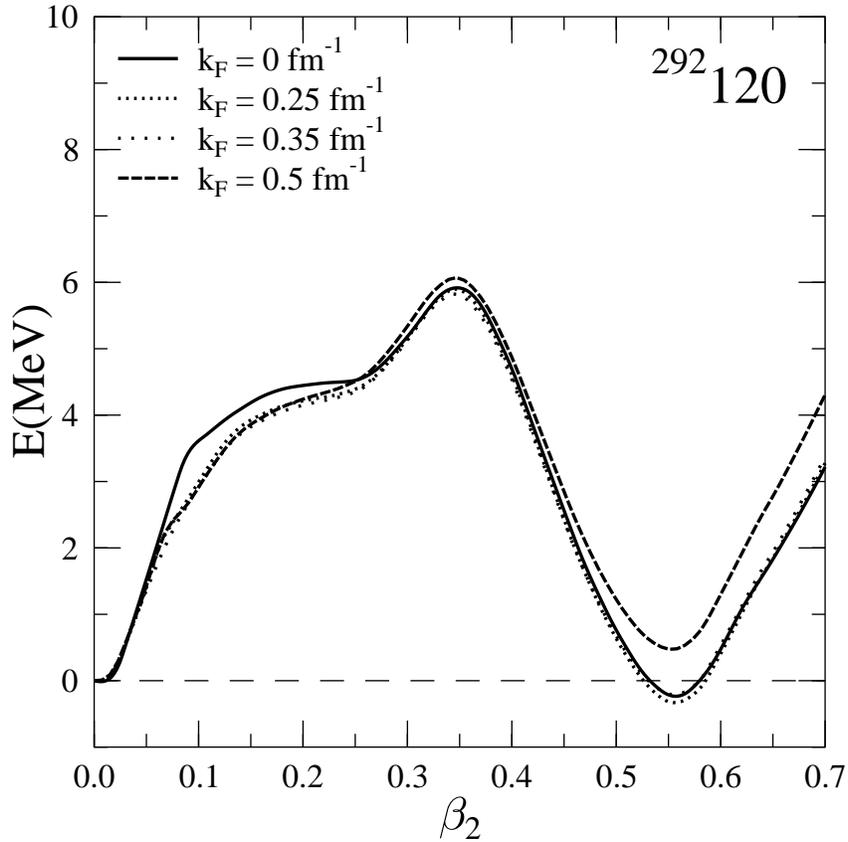}}
\caption{Fission barrier of the superheavy nucleus  $^{292}120$ for various electron densities in spheroidal configurations.}
\label{120_172}
\end{figure*}
Fig. \ref{120_172} demonstrates the influence of the electron background on the superheavy
nucleus $^{292}$120. In this case, the electron background leads to a slight decrease
of the width and height of the inner barrier calculated with the spheroidal electron cloud. But the height of the second minimum has increased by about 0.7 MeV. Note, however, that 
the inclusion of reflection-asymmetric degrees of freedom in these superheavy
systems make the second barrier vanish \cite{shebarrier}. It is interesting that the
potential energy surface is affected in a different way compared to the case of Plutonium.
The reasons are related to the fact that the barrier starts out from
a spherical minimum, and that the charge number in this nucleus is much larger so that
the total contribution of the Coulomb energy to the total energy is strongly
increased compared to plutonium. 

\begin{figure*}[t]
\centerline{\epsfxsize=12cm \epsfbox{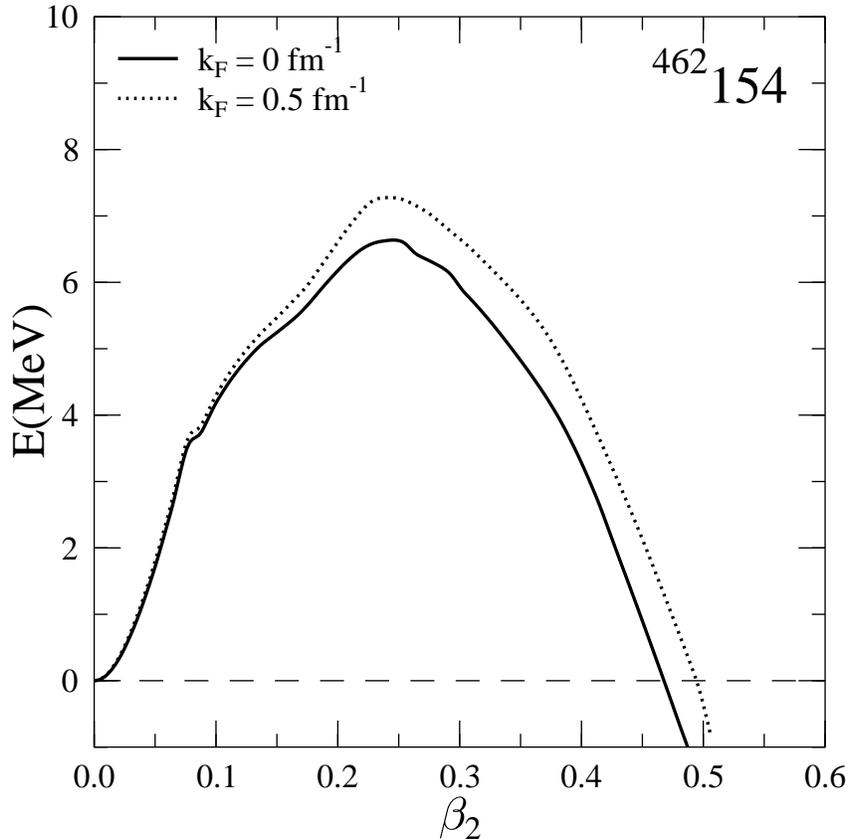}}
\caption{Fission barrier of the superheavy nucleus  $^{462}154$ without electrons and for $k_F = 0.5$~fm$^{-1}$}
\label{154}
\end{figure*}

We can speculate about even heavier systems. In Ref. \cite{shell-corrections}, the shell corrections
for super- and hyperheavy nuclei were calculated within the framework of self-consistent
mean-field models. It was found that the familiar concept of strongly-pronounced magic numbers dissolves due to the high level-density, and rather a broad region of shell-stabilization can be found.
We choose the hyperheavy nucleus $^{462}154$, which is located in such a region of shell-stabilization
(for the RMF force NL3), as an example. Its fission barrier can be seen in Fig. \ref{154}.
The strong shell corrections lead to a spherical minimum and to a barrier which is which is about 6.5 MeV
high, and almost 1 MeV larger in the presence of electrons. 

A word of caution is in order. For both this hyperheavy nucleus and the superheavy system $^{292}120$
discussed before,
the calculation of the axial barrier gives us only partial information on the principal impact of
the electron background on the fission barrier.
The actual fission path, however, might deviate
quite strongly from this path, and for these heavy systems fissioning through the triaxial
plane is rather probable.

\subsubsection{Fission Q values}
\begin{figure*}[t]
\centerline{\epsfxsize=10cm \epsfbox{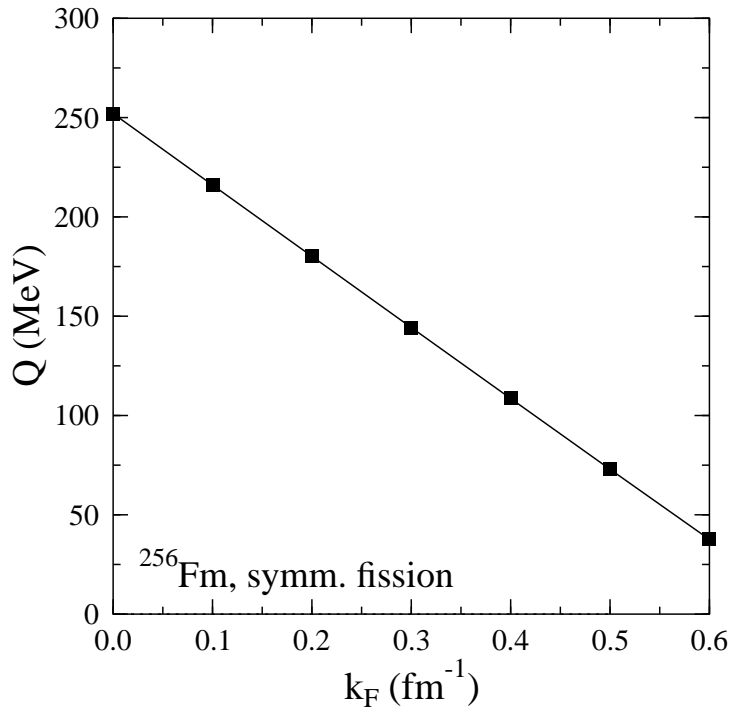}}
\caption{Q value for symmetric fission of $^{256}$Fm as a function of the electron Fermi momentum $k_F$.}
\label{q_fission}
\end{figure*}

As discussed in Section \ref{WSsph}, due to the screening of the electrons the 
Coulomb energy of nuclei is lowered as compared with the vacuum case.
This will lead to the reduction of the Q value for spontaneous fission, in full analogy
to the case of $\alpha$ decay. Fig. \ref{q_fission} shows the Q value for symmetric fission of $^{256}$Fm as a function of the electron Fermi momentum. One can clearly see this trend and we think it will be present
for asymmetric fission too. To evaluate the fission probability one should
perform detailed calculations of the fission barrier for the case of a two-center nuclear
shape. This work is in progress now. At this stage we can only point out that due to the screening
of the Coulomb potential the fission barrier is altered and the Q value will decrease. We expect that the
latter trend is more important so that the net effect is suppression
of spontaneous fission. This follows also from simple estimates based on the liquid-drop
model. Indeed, let us consider the fissility parameter
\begin{equation} 
\frac{Z^2}{A} = \frac{2 a_S}{a_C},
\end{equation}
where $a_S \approx 18 ~MeV$
and $a_C \approx 0.72 ~MeV$ are, respectively the surface and Coulomb coefficients in the
Weizs\"acker formula. If we take into account the electron screening, the Coulomb energy
is reduced by the factor $c$, Eq. (\ref{e_coulomb_theta}). Therefore the modified fissility
parameter is $\frac{Z^2}{A} = \frac{50}{c}$. Since $c(x) < 1$, this means that in the presence of electrons
the region of spontaneous fission moves to heavier nuclei. For instance, for $k_F = 0.25 fm^{-1}$,
the fissility parameter is 80 instead of 50 in vacuum.

\section{Conclusions and Outlook}
\label{conclusions}
We have studied the properties of atomic nuclei
embedded in
an electron gas as occuring in, e.g., neutron star
crusts and supernova explosions. Nuclear structure calculations have been performed
within the relativistic mean-field approach employing the force NL3.
The calculations have been performed within the Wigner Seitz cell approximation
in coordinate-space. 
A fermi-type distribution with a smooth
surface has been employed in order to get rid of artifacts stemming from the discrete grid spacing.
For spherical systems, a spherical Wigner-Seitz cell has been used.
We have discussed 
an implementation of  spheroidal Wigner-Seitz cells for axially-deformed nuclei.
We employ the criterion that in addition to the total charge also the total electric quadrupole moment of a cell should vanish. This leads to an
iterarively self-adjusting Wigner-Seitz cell for each given deformation of the nucleus.

The presence of electrons leads to effects
which have rather macroscopic character. 
We did not find any significant
changes in the single-particle
levels regarding their relative positions. Since the electron background is constant over the nuclear volume -- by construction --
it leads predominantly to a constant downward shift of the proton potential.

The electron gas alters the stability condition for $\beta$ decay and
shifts the stability line to more neutron-rich nuclei. Electron capture becomes
a relevant process, the transformation of protons into neutrons favors large
isospin in nuclei.
Furthermore, the two-proton dripline
is shifted to more proton-rich nuclei since the protons gain additional binding due to the 
attractive interaction with the electrons.
The neutron drip line is not altered by the presence of the electrons since the neutron
single-particle potential remains (to a very good approximation) unaffected.

We have found that in a dense electron background decay modes such as $\alpha$-decay and spontaneous
fission are suppressed. Our calculations show that the $\alpha$ decay half lives decrease
a a function of the electron Fermi momentum $k_F$ until they get stabilized for larger $k_F$.
A general trend is that the $Q$ values of those decay modes decrease with increasing electron
density.

We have also calculated the potential energy surface as a function of the electron
Fermi momentum within the framework of the self-adjusting axial Wigner Seitz
cell. We have found that only for very large electron Fermi momenta 
a change is occuring for the inner axial barrier and the energetic position of the isomeric
state. This could hint again at stabilization of the fission mode for 
high electron densities.

Overall, the electron gas broadens the part of the nuclear chart lying between the proton drip line
and the valley of $\beta$- stability. Stabilizing effects with respect to $\alpha$ decay and spontaneous fission
occur for large electron
Fermi momenta. Hence, in
extreme astrophysical environments, the production of very exotic and superheavy nuclei 
could become possible. This might happen during the r-process (rapid neutron capture)
and the rp-process (rapid proton capture)
when
the electron background prevents the heavy (and superheavy) nuclei from fast decay
by spontaneous fission or alpha decay.
As a consequence, the nuclei, which otherwise would be unstable, can provide a bridge
to the island of superheavy elements. We are planning to study this possibility
in the future.
If long-lived or even stable superheavy nuclei exist, they
could be created in such an environment and later ejected into  space. Then one
can try to search for such superheavy nuclei in cosmic rays.
Another natural step in our investigations is to add free neutrons and 
study  nuclei in the environment of free electrons and neutrons. Work in this direction is in progress.

\section*{Acknowledgement}
\label{acknowledge}
Fruitful discussions with J. A. Maruhn, S. Reddy, P.--G. Reinhard, N. Sandulescu, L. Satarov, J. Schaffner-Bielich and S. Schramm are gratefully
acknowledged. This work was supported in part by the Gesellschaft f\"ur Schwerionenforscung
(Germany) and by grants RFFR-02-04013 and NS-8756.2006.2 (Russia).

\section*{Appendix}
\label{appendix}
\subsection{Role of Coulomb exchange}
In relativistic models such as used in this study,  the Coulomb interaction is usually
calculated within the Hartree approximation, i.e. 
\begin{equation}
{\cal E }_{Coulomb}^{dir} 
 = e^2 \frac{1}{2} \int \int d^3 r d^3 r^{'} 
\frac{\rho_p(\vec{r})\rho_p(\vec{r}^{'})}{|\vec{r}-\vec{r}^{'}|}
\end{equation}
where $\rho_p(\vec{r})$ is the proton density. This relates to the fact that the
RMF model traditionally is formulated as a Hartree theory, and no explicit exchange
terms in the effective nucleon-nucleon interaction are taken into account.
In contrast to effective field theories with contact interactions, exchange terms involving
finite-range meson fields are computationally expensive, and so far no significant
improvement over Hartree-like formulations has been achieved. 
The role of 4-fermion interactions has been studied in Ref. \cite{4fermion}
for the point-coupling variant of the RMF model, the RMF-PC model. It has been found that,
while the formal structure of the model remains the same and only a redefinition
of 4-fermion coupling constants occurs, the interpretation of the various
terms becomes quite different. For higher-order point-coupling terms and derivative
terms, however, the Fierz transformations \cite{fierz} yield a large number of terms which
do not improve the model \cite{cornelius}.

In the Skyrme-Hartree-Fock approach, the Coulomb exchange term is usally
included in Slater approximation \cite{slater}:
\begin{equation}
{\cal E }_{Coulomb}^{ex} 
 = \frac{3}{4} e^2 \Big(\frac{3}{4}\Big)^{1/3} \int d^3 r [\rho_p(\vec{r})]^{4/3}
\end{equation}
In our calculations, we replace theproton density by the absolute value of the charge density, i.e., $\rho_p(\vec{r})$ by $|\rho_{ch}(\vec{r})| = |
\rho_p(\vec{r}) + \rho_e(\vec{r})|$.

\begin{figure*}[t]
\centerline{\epsfxsize=16cm \epsfbox{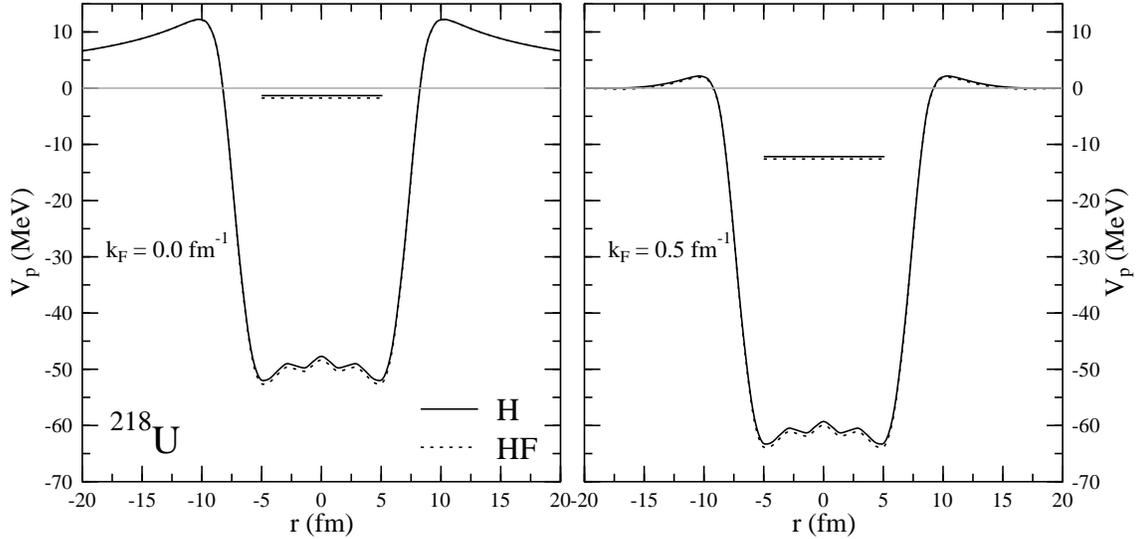}}
\caption{The proton single-particle potential for $^{218}$U in the absence
of electrons (left) and for $k_F = 0.5 fm^{-1}$ (right) involving only
the direct Coulomb term (H) and involving both direct term and exchange term
in Slater approximation (HF). Also shown are the proton fermi energies for both cases.}
\label{coul}
\end{figure*}

For nuclei close to stability, it can be shown that the Coulomb exchange effect can be absorbed
by a refit of the mean-field parameters \cite{coulex}. It remains to be seen, however,
if this still holds true for exotic nuclei and nuclei close to the proton drip-line,
as well as nuclear systems studied here.
As demonstrated in Fig. \ref{coul}, the Coulomb exchange interaction both in vacuum
and at high electron chemical potential
yields a rather small contribution to the potential, and 
Fermi energies for
the two cases differ by 0.4 MeV only. Thus the properties of nuclei in an electron gas are not
sensitive to the exclusion or inclusion of (Slater-) Coulomb exchange in the
calculation.

\end{document}